\documentclass[aps,prl,twocolumn,floats,showpacs,superscriptaddress]{revtex4-1}
\usepackage{graphicx,epsfig,psfrag}
\usepackage{times}
\usepackage{graphics,dcolumn,bm,float}
\usepackage{amssymb,amsmath,rotate,color}
\usepackage[title,titletoc,toc]{appendix}
\usepackage{graphicx}

\begin{document}
\unitlength 1 cm
\newcommand{\be}{\begin{equation}}
\newcommand{\ee}{\end{equation}}
\newcommand{\ba}{\begin{align}}
\newcommand{\ea}{\end{align}}
\newcommand{\bearr}{\begin{eqnarray}}
\newcommand{\eearr}{\end{eqnarray}}
\newcommand{\nn}{\nonumber}
\newcommand{\la}{\langle}
\newcommand{\ra}{\rangle}
\newcommand{\cd}{c^\dagger}
\newcommand{\vd}{v^\dagger}
\newcommand{\ad}{a^\dagger}
\newcommand{\bd}{b^\dagger}
\newcommand{\tk}{{\tilde{k}}}
\newcommand{\tp}{{\tilde{p}}}
\newcommand{\tq}{{\tilde{q}}}
\newcommand{\eps}{\varepsilon}
\newcommand{\vk}{\vec k}
\newcommand{\vp}{\vec p}
\newcommand{\vq}{\vec q}
\newcommand{\vkp}{\vec {k'}}
\newcommand{\vpp}{\vec {p'}}
\newcommand{\vqp}{\vec {q'}}
\newcommand{\bk}{{\bf k}}
\newcommand{\bp}{{\bf p}}
\newcommand{\bq}{{\bf q}}
\newcommand{\br}{{\bf r}}
\newcommand{\bR}{{\bf R}}
\newcommand{\up}{\uparrow}
\newcommand{\down}{\downarrow}
\newcommand{\fns}{\footnotesize}
\newcommand{\ns}{\normalsize}
\newcommand{\cdag}{c^{\dagger}}

\newcommand{\sx}{\sigma^x}
\newcommand{\sy}{\sigma^y}
\newcommand{\sz}{\sigma^z}
\newcommand{\nt}{n^\theta}

\title{Semiconductor of spinons: from Ising band insulator to orthogonal band insulator}
\author{T. Farajollahpour}
\affiliation{Department of Physics, Azarbaijan Shahid Madani
University, 53714-161, Tabriz, Iran}
\affiliation{Department of Physics, Sharif University of Technology, Tehran 11155-9161, Iran}

\author{S. A. Jafari}
\email{akbar.jafari@gmail.com}
\affiliation{Department of Physics, Sharif University of Technology, Tehran 11155-9161, Iran}
\affiliation{Center of excellence for Complex Systems and Condensed Matter (CSCM), Sharif University of Technology, Tehran 1458889694, Iran}

\begin{abstract}

 Within the ionic Hubbard model, electron correlations transmute the single-particle gap of a band insulator into a Mott gap in the strong correlation limit. However understanding the nature of possible phases in between these two extreme insulating phases remains an outstanding challenge. We find two strongly correlated insulating phases in between the above extremes: (i) The insulating phase just before the Mott phase can be viewed as gapping a non-Fermi liquid state of spinons through staggered ionic potential. The quasi-particles of underlying spinons are orthogonal to physical electrons and hence they do not couple to photoemission probes, giving rise to "ARPES-dark" state due to which the ARPES gap will be larger than optical and thermal gap.  (ii) The correlated insulating phase just after the normal band insulator corresponds to the ordered phase of slave Ising spins (Ising insulator) where charge configuration is controlled by an underlying Ising variable which indirectly couples to external magnetic field and hence gives rise to additional temperature and field dependence in semiconducting properties. In the absence of tunability for the Hubbard $U$, such a temperature and field dependence can be conveniently employed to achieve further control on the transport properties of Ising-based semiconductors. The rare earth monochalcogenide semiconductors where the magneto-resistance is anomalously large can be a candidate system for the ordered phase of Ising variable where pairs of charge bosons are condensed in the background. Combining present results with our previous dynamical mean field theory study, we argue that the present picture holds if the ionic potential is strong enough to survive the downward renormalization of the ionic potential caused by Hubbard $U$.
\end{abstract}
\maketitle

\section{INTRODUCTION}
Electron conduction in periodic structures can cease for two reasons. The simplest
is to couple single-particle states across a reduced Brillouin zone by an 
off-diagonal matrix elements due to reduction in the periodicity. However the
second and more exciting way is to introduce strong electron correlations where
due to Coulomb interactions, as suggested by N. Mott, electron conduction
in an otherwise conducting state is interrupted~\cite{MottBook}.
This may seem to suggest that strong correlation has its most dramatic effect
on metals by transforming them into many-body Mott insulators. 
The canonical model within which the metal-to-insulator transition (MIT) 
problem is investigated is the Hubbard model~\cite{Fujimori}.
Efforts to understand the nature of MIT has lead to many 
technical~\cite{Gutzwiller,MuthukumarGros,Metzner,Senechal,Adibi,Vladimir,OvchinikovBook,Sorella,Imada,Kotliar,Tohyama,Dagotto}
and conceptual~\cite{LeeWenNagaosa,Florens2004,Sigrist,HasanSlaveSpin,Zaanen} developments
providing clues into possible mechanisms of non-Fermi liquid formation.

But even more challenging question is what happens when both mechanisms of gap
formation are simultaneously present, i.e. what are the properties of strongly correlated band insulators
or semiconductors?
Let us formalize the problem as follows: Imagine a staggered potential of strength $\Delta$
(the ionic potential) that can gap out the parent metallic state and sets the scale of 
the single-particle gap. 
When the Hubbard interaction $U$ is turned on in such an already
gapped state (band insulator) an interesting competition between the Hubbard $U$ and
the staggered potential $\Delta$ sets in. This is the simplest model addressing the
competition between a "many-body" gap parameter $U$ and a "single-particle" gap
parameter $\Delta$ which is called ionic Hubbard model~\cite{Egami}. The band insulating
state at $U=0$ is adiabatically connected to the insulating state at non-zero but small values of $U\ll \Delta$
as the effect of weak Hubbard $U$ is to renormalize the parent metallic state
on top of which the ionic potential creates a band insulator (BI) state. 
In the opposite limit of strong
correlations $U\gg \Delta$ one gets a Mott insulating (MI) state. Although these
two extreme limits both represent insulating states, the origin of gap in the former case is
a simple one-particle scattering, while in the later case the gap has a many-body character
arising from projection of doubly occupied configurations.

The nature of possible state(s) between the above two extreme insulating states has
been the subject of debates in the past decade. In one dimension Fabrizio and coworkers~\cite{Fabrizio}
find an ordered state. In two dimensions Hafez-Torbati and coworkers find orientational and
bond ordering phase in between the Mott and band insulators~\cite{Hafez2015}.
The topologically non-trivial variant of the model was considered by
Prychynenko and coworkers~\cite{Huber} who find for topologically trivial situation
two spin density wave states are sandwiched between the band and Mott insulating states.
In the limit of infinite dimensions however, Garg and coworkers using the dynamical mean field theory found that
the competition between the tendency of the ionic potential $\Delta$ and the Hubbard term $U$
gives rise to a metallic state~\cite{Garg}. Within a perturbative
continuous unitary transformation one finds a metallic state when
the Hubbard $U$ and the ionic potential $\Delta$ are comparable~\cite{Hafez2009}. Similar 
result were obtained in two dimensions~\cite{Baudim}. The method of dynamical mean field theory
was also applied to study the quantum phase transitions of the ionic Hubbard model
on the honeycomb lattice.
Starting from massive Dirac fermions
on the honeycomb lattice the competition between $U$ and the single-particle gap parameter $\Delta$
(known as mass term when it comes to Dirac fermions) gives rise to massless Dirac
fermions~\cite{Ebrahimkhas}. A recent strong coupling expansion gives a
quantum critical semi-metallic state~\cite{Adibi}.

When electron correlations in a conductor are not strong enough to transform the metallic
state to a Mott insulator, they give rise to possible non-Fermi liquid states. 
From this point of view one may now turn on the ionic potential $\Delta$ and ask the following question:
How does this staggered potential interfere with possible non-Fermi liquid state 
of the parent metal? Can the ionic potential gap out a non-Fermi liquid state?
With this motivation, let us summarize one of the simplest mechanisms of 
creating a non-Fermi liquid state, and then add the ionic potential $\Delta$ to it
in a self-consistent way.
This is the question on which we will be focused in this paper.

Recently Nandkishore and coworkers have argued that starting from a metallic state,
as one increases the Hubbard $U$ beyond $U_\perp$ the Fermi liquid (FL) undergoes a
phase transition to an exotic non-FL state termed orthogonal metal (OM). Upon further increase
in the interaction strength beyond $U_c$ the system becomes a Mott insulator~\cite{OMPhys,Nandkishore}.
OM is an interesting -- and perhaps the simplest -- non-FL state for $U_\perp<U<U_{\rm Mott}$
that separates a FL from a Mott insulator. 
In the FL phase the rotor variable is ordered, i.e. the phase variable
has small fluctuations meaning that the electric charge has large fluctuations.
When the rotor variable disorders, earlier interpretation would assume that wild
fluctuations of the rotor field that is responsible for vanishing of the rotor variable
corresponds to freezing of charge fluctuations and hence making the system a Mott insulator~\cite{Florens2004}.
However, Nandkishore and coworkers
argued that even if the rotor variable is disordered, i.e. $\la e^{i\theta}\ra=0$
for $U>U_b$
meaning that the single boson $b\sim e^{i\theta}$ is not condensed, a two-boson combination
can still be condensed, $\la bb\ra\ne 0$. This gives a new chance to charge fluctuations to
survive in the form of an Ising variable which then can be naturally captured within
a slave spin approach~\cite{Sigrist} where the condensation of two-boson combination is
reflected in a non-zero Ising order parameter for $U_b<U<U_\perp$. 
This phase can be dubbed Ising metal. The disordered phase of such an Ising
variable for $U_\perp<U<U_{\rm Mott}$ will correspond to OM state where although quasiparticle
weight corresponding to physical electrons is zero, its transport behaviour is metallic.
To see this, the following simple and powerful argument is due to Nandkishore and coworkers:
Within the slave spin representation the physical electron is represented
as $c^\dagger_{\sigma}=f^\dagger_{\sigma} \tau^x$. The $U(1)$ transformation representing 
the conservation of the electric (Noether) charge can only be incorporated into
the spinon $f^\dagger_\sigma$, simply because the Pauli matrix $\tau^x$ is purely real.
Therefore the $f$ spinon inherits the whole charge from
the parent electron and after the Ising disordering transition at $U_\perp$ the 
spinons will continue to display metallic transport properties despite that the 
quasiparticle weight of the physical electrons has been already lost at $U_\perp$ as
it is proportional to Ising magnetization~\cite{Nandkishore}.
If instead of a metallic state at $U=0$, one starts with a semi-metallic state
(such as graphene) the state corresponding to disordering transition Ising pseudo-spins
will be an orthogonal semi-metal~\cite{Chinese}. For our purpose in the present
paper, we would like to see that, both the orthogonal metal phase for $U_\perp<U<U_{\rm Mott}$
and the Ising metal (IM) for $U_b<U<U_\perp$ are equally interesting when one gaps out the
underlying metallic state.

\begin{figure}[t]
\includegraphics[width=0.9\linewidth]{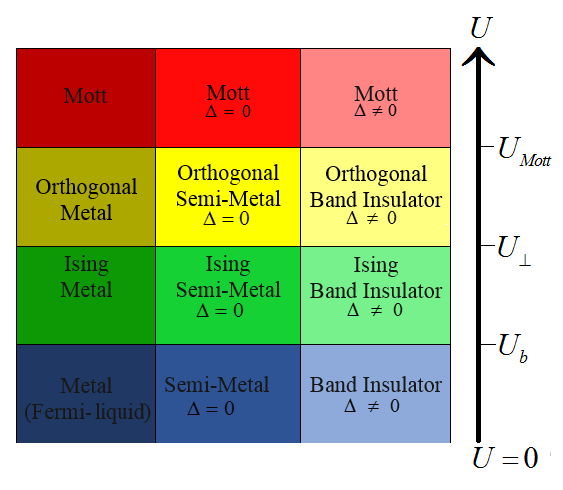}
\caption{ (Color online) The schematic representation of the effect of
correlation on three starting states. The non-interacting
state from which we start is drawn in the bottom row. Left column
corresponds to metallic state, middle column stands for semi-metallic state
and the right column denotes the band insulating state. All columns
at very large $U$ end in the Mott insulating states (top row).
The middle row is the corresponding "orthogonal" phase where the
transport is controlled by the spinons that inherit the charge of
the original electrons.
}
\label{schematic.fig}
\end{figure}

Now let us turn on the ionic potential $\Delta$ that locally couples to the 
electric charge as $(-1)^i \Delta c^\dagger_{i\sigma} c_{i\sigma}$ at every site $i$ 
of the lattice. The Ising term $\tau^x$ is "eaten up" by the $U(1)$ invariance of this term
and the spinon density directly couple to the ionic potential as $(-1)^i\Delta f^\dagger_{i\sigma}f_{i\sigma}$.
This is because spinons carry the whole charge and hence their density couples to 
external electrostatic potential (including even random potentials).
The above term clearly gaps out both the IM and the OM conducting state of spinons and creates a 
band insulator of Ising metal and spinon metal. These phases can be called 
Ising band insulator (IBI) and orthogonal band insulator (OBI), respectively as
they are born out of an underlying IM and OM states. When the (thermal) gap of the OBI is small enough 
to comply with semiconducting gaps we will have an orthogonal semiconductor, i.e.
a semiconductor of spinons that has been separated from an Ising semiconductor.

Our discussion is schematically summarized in Fig.~\ref{schematic.fig}
where the notion of Ising and orthogonal state before a Mott state has been illustrated
for three weak coupling states: metal, semimetal, and band insulator. 
The weak coupling states at the bottom row (blue) turn into their Ising counterpart (green)
by increasing $U$ beyond $U_b$ where at which the single boson condensate (rotor order parameter)
vanishes. In the Ising (green) phase, still the physical electrons are governing the
transport properties, but the charge condensate is survived as an Ising order which
eventually vanishes at $U_\perp$. From this point the orthogonal (yellow) phase starts.
By further increasing the Hubbard $U$ beyond $U_{\rm Mott}$ even the residual interaction 
between the spinons of the orthogonal phase become so strong that renders the system Mott insulating.

Therefore the intuitive picture that emerges for the ionic Hubbard model (right column) 
is as follows: At small values of Hubbard $U$ (blue region)
the gapped state is adiabatically connected to a band insulator. When $U$ crosses
$U_b$ at which the single bosons "quantum evaporate" from the condensate but the
boson pairs form a condensate, we will have IBI where the charge variable is
Ising-like and its characteristic Ising-like properties is expected to give rise to
unusual semiconducting properties (green region). By further increasing $U$ up to $U_\perp$ where
the Ising order parameter of IBI vanishes, the OBI phase (yellow region) starts which is eventually
gapped by prohibition of doubly occupancy at $U_{\rm Mott}$.

In the OBI phase for $U_{\perp}<U<U_{\rm Mott}$ (yellow region) the disordering of the underlying
slaved Ising variables leaves the states in bottom of conduction and top of valence
bands inaccessible to ARPES, while accessible to any probe coupling to the electric current (charge),
such as optical conductivity, cyclotron resonance and thermal probes. This plays a significant
role in experimental discrimination of the orthogonal insulators (semiconductors) 
from their Ising (green) or normal (blue) relatives. The bandwidth and hence the effective mass of the
correlated semiconducting phase IBI is controlled by Ising order parameter which will then possess 
Ising-like temperature dependencies and hence e.g. cyclotron frequency will acquire
Ising-like temperature dependence, etc; while the OBI phase is characterized by wider ARPES gap 
compared to thermal gap. 

The results of this paper apply to a quite general correlated insulators on any
two-dimensional lattice. 
However in this paper we focus on honeycomb system where the parent
band insulator is described by massive Dirac electrons~\cite{GappedEpitaxialGraphene}.
At the end we contrast the above possibilities with our previous 
dynamical mean field study of the ionic Hubbard model on honeycomb lattice~\cite{Ebrahimkhas} 
to discuss plethora of Ising and orthogonal phases that maybe conducting or insulating.
This will shed a new light: To get a chance to realize IBI the ionic potential $\Delta$ must
be large enough. To realize OBI, the ionic potential must be even larger than what is
required to realize IBI.
For very small values of $\Delta$, only the Ising and orthogonal phase of massless
Dirac fermions can be realized which corresponds to a orthogonal semi-metal.

This paper is organized as follows: In section II after introducing the
ionic Hubbard model, we review the slave-rotor method and customize it for
the ionic Hubbard model. In section III we adopt the version of slave-spin method
employed in~\cite{Sigrist}. Based on symmetry principles we discuss under
what circumstances the Lagrange multiplier implementing the constraint between
Ising pseudo-spin and spinon degrees of freedom vanishes. In section IV we discuss
the results and summarize the findings at the end. 

\section{Ionic Hubbard model and slave rotor method}
We are interested in the phase diagram of the Hubbard model augmented by
a staggered ionic potential of strength $\Delta$ as follows:
\bearr
&&H=H_0 + \frac{U}{2}\sum_i (n_{i\up}+ n_{i\down}-1)^2 \label{hubbard.eqn}\\
   &&H_0= -t\sum_{\langle i,j\rangle \sigma} c^\dagger_{i\sigma} c_{j\sigma}
   +\Delta\sum_{i\sigma} (-1)^i  n_{i\sigma}-\mu\sum_{i\sigma} n_{i\sigma}
\eearr
where $c^\dagger_{i\sigma}$ creates an electron at a localized orbital in site $i$
with spin $\sigma$, $n_{i\sigma}=c^\dagger_{i\sigma}c_{i\sigma}$ is the occupation
number, $U$ is the on-site Hubbard repulsion, $t$ is the hopping amplitude
which will be set as the unit of energy through out the paper, $\Delta$ is the ionic
potential, and $\mu$ is the chemical potential that at half-filling in the present
representation turns out to be $\mu=0$.

\subsection{Slave rotor method}
The slave rotor formulation is one of the slave particle family methods employed
in studying the Hubbard model that provides a very economical representation of the
charge state of an orbital in terms of a rotor variable conjugate to an
angular momentum operator locked to the charge~\cite{Florens2004}.
This method can also be applied to the study of Anderson impurity problem~\cite{Florens2002}.
In this approach the local Hilbert space is represented by a direct product
of the Hilbert space of a fermion carrying the spin index (the so called
spinon) and a rotor that controls the charge state of the system as
$\left| \psi  \right\rangle =\left|
{{\psi }_{f}} \right\rangle \left| {{\psi }_{\theta }} \right\rangle $.
In terms of operator creating a particle from its vacuum the above
equation can be represented as,
\begin{eqnarray}
\hat{c}_{i\sigma }^{\dagger }=f_{i\sigma }^{\dagger }{{e}^{-i{{\theta }_{i}}}},
\label{slaverotor.eqn}
\end{eqnarray}
where $f^\dagger_{i\sigma}$ is the spinon creation operator and $\theta_i$ is the
rotor variable at site $i$.
The physical Hilbert space of the electron in terms of the spinons and rotors is constructed 
from the following states:
\begin{align}
  & \left| 0 \right\rangle \equiv \left| 0 \right\rangle_f \left| -1 \right\rangle_\theta  \nn\\
 & \left| \up  \right\rangle \equiv \left| \up  \right\rangle_f \left| 0 \right\rangle_\theta   \nn\\
 & \left| \down  \right\rangle \equiv \left| \down  \right\rangle_f \left| 0 \right\rangle_\theta \nn\\
 & \left| \up \down \right\rangle \equiv \left| \up \down  \right\rangle_f \left| 1 \right\rangle_\theta
   \label{states.eqn}
\end{align}
where $|\ra_\theta$ represents a state in the rotor space, and $|\ra_f$ represents
states in the spinon Hilbert space. As can be seen in the above representation,
state such as $|\up\ra_f |-1\ra_\theta$ containing one $\up$-spin spinon and corresponding
to angular momentum eigen state of $-1$ does not correspond to any physical state. Such redundancy
is a characteristic of auxiliary particle methods where the physical Hilbert space is enormously
enlarge. What we gain in the enlarged Hilbert space is the gauge freedom. But the physically
sensible states are obtained from those in the enlarged space by projecting them to the
physical space. In the present case such a projection amounts to the constraint,
\be
   \sum\limits_{\sigma }{f_{i\sigma }^{\dagger } {{f}_{i\sigma }}}=L_i+1
   \label{gheid.eqn}
\ee
that locks the particle number to angular momentum.
It can be seen that all of the states in Eq.~\eqref{states.eqn} satisfy this
constraint. If the above constraint can be implemented exactly, then either
of the representation in terms of original electrons or in terms of spinons and
rotors will describe the same physics. But the full-fledged implementation of the above
projection requires to take a complete care of the fluctuations of the internal
gauge fiedls that glue spinons and rotor fields. In the present
work we treat the constraint in the mean field via a space- and time-independent
Lagrange multiplier. This is known as the slave rotor mean field
approximation~\cite{Florens2004,Paramekanti}.

Let us proceed with representing the electron operators in terms of spinons and rotors,
Eq.\eqref{slaverotor.eqn},
\begin{align}
   &H=-\sum\limits_{\left\langle ij \right\rangle ,\sigma }{f_{i\sigma }^{\dagger }
{{f}_{j\sigma }}{{e}^{-i{{\theta }_{ij}}}} +{\rm h.c.}}+\Delta \sum\limits_{i}{f_{i}^{\dagger }
{{f}_{i}}}-\Delta \sum\limits_{j}{f_{j}^{\dagger }{{f}_{j}}}\notag\\
   &+\frac{U}{2}\sum\limits_{i}{L_{i}^{2}-\mu \sum\limits_{i\sigma }{f_{i\sigma }
^{\dagger }{{f}_{i\sigma }} }} + \lambda\sum_i \left(f^\dagger_{i\sigma} f_{i\sigma}-L-1 \right),
\end{align}
where the hopping amplitude $t$ of original electrons is set as the unit of
energy, $t=1$, $\theta_{ij}=\theta_i-\theta_j$, the Hubbard term of Eq.~\eqref{hubbard.eqn}
has been transformed with the aid of the constraint~\eqref{gheid.eqn},
$\mu$ is the chemical potential that becomes zero at half-filling which is our focus
in this paper, and $\lambda$ is a position-independent Lagrange multiplier that implements
the constraint~\eqref{gheid.eqn} on average. The merit of representation in terms of
auxiliary rotor variables is that the quartic interaction between the fermions
$U/2{{\sum\nolimits_{i\sigma }{\left( n_{i\sigma }-1 \right)}}^{2}}$ is replaced by a simple
rotor kinetic energy $U {{L}^{2}}/2$ at every site, 
where the angular momentum $L= -i{{\partial }_{\theta }}$
is a variable that is associated with a $O(2)$ quantum rotor $\theta$.

Apart from the kinetic term that involves spinons and rotors on neighbouring sites,
the above Hamiltonian is decoupled into rotor ($\theta$-only), and spinon ($f$-only) terms.
As for the first term we introduce mean field variables,
\bearr
&&\chi _\theta =\left\la e^{i\theta_i} e^{-i\theta_j} \right\ra_\theta, \label{chitheta.eqn}\\
&&\chi_f= \left\la \sum\nolimits_\sigma f_{\sigma i}^\dagger{f}_{\sigma j} \right\ra_f \label{chif.eqn},
\eearr
to decouple the kinetic term, which eventually gives decoupled rotor, $H_\theta$ and spinon
$H_f$ Hamiltonians,
\begin{align}
   &H_f=-\sum\limits_{\la ij \ra ,\sigma}f_{i\sigma }^\dagger f_{j\sigma }\chi_\theta + {\rm h.c}
   +(\lambda-\mu)\sum_{j\sigma} f^\dagger_{j\sigma} f_{j\sigma}  \nn\\
   &+\Delta \sum_{i\in A} f_{i\sigma}^\dagger f_{i\sigma}-\Delta\sum_{j\in B} f_{j\sigma}^\dagger f_{j\sigma},
   \label{spinon.eqn}
\end{align}
and
\be
  H_\theta=\sum_{\la i,j\ra} \chi_f e^{i\theta_i-i\theta_j}+h.c.
  +\sum_j\left( \frac{U}{2} L_j^2 - \lambda L_j\right),
  \label{rotor.eqn}
\ee
where $A$ and $B$ are the two sub-lattices on the honeycomb lattice.
In the above equation, solution of the $H_f$ requires a knowledge of $\chi_\theta$
which according to Eq.~\eqref{chitheta.eqn} can only be calculated after having diagonalized
the rotor sector $H_\theta$. The later itself depends on unknown quantity $\chi_f$ that according to
Eq.~\eqref{chif.eqn} can be obtained from the $H_f$ Hamiltonian. This provides a self-consistency
loop, i.e. the rotor and the spinon sectors talk to each other via the mean-field self-consistency
equations,~\eqref{chitheta.eqn} and~\eqref{chif.eqn}.

It is interesting to note that in Eq.~\eqref{spinon.eqn} the ionic potential $\Delta$, being
a local potential couples only to the spinon density. This considerably simplifies the analysis
of the problem. Indeed in the absence of ionic potential $\Delta$, the  spinon sector
would have been described by spinon hopping Hamiltonian with renormalized hopping amplitudes
whose reduced kinetic energy is encoded in $\chi_\theta$ which is 
self-consistently determined by the rotor sector. 
When the ionic potential is turned on, the spinon sector describes spinons
hopping with renormalized hopping parameters $\chi_\theta$, plus an additional Bragg reflection
due to doubling of the unit cell that always gaps out the spinon sector and the
spectrum in the spinon-sector becomes,
\begin{eqnarray}
\eps_f(\vk)=\pm \sqrt{\chi_\theta^2|\phi(\vk)|^2+\Delta^2}
\end{eqnarray}
where $\phi(\vk)=1+e^{i\vk.\vec a_1}+e^{i\vk . \vec a_2}$ with $\vec a_1$ and $\vec a_2$
unit being translation vectors of the honeycomb lattice.

Now let us turn to determination of chemical potential $\mu$ and Lagrange multiplier field
$\lambda$. Since we are interested in the competition between $U$ and $\Delta$, we stay at half-filling
where even in the $U=0$ limit the system is described by a simple band insulator.
The particle-hole symmetry of the original Hamiltonian in terms of physical electrons
simply implies that at half-filling $\mu=0$. Since we want to fix the average occupation
at $n=1$, the constraint~\eqref{gheid.eqn} implies that on average one should have $\la L_j \ra=0$
for every lattice site $j$. Now imagine that in a rotor Hilbert space corresponding to a given
total angular momentum $\ell$, we construct the local angular momentum operator 
$UL_j^2/2-\lambda L_j$
that would take $2\ell+1$ diagonal values $U\left( n^\theta_j \right)^2-\lambda n^\theta_j$, where
$n^\theta_j=-\ell \ldots \ell$ represents all possible values of the magnetic quantum number.
Obviously any non-zero value of $\lambda$ breaks the symmetry between the states corresponding to
positive and negative values of $n^\theta_j$ and therefore makes the expectation value $\la L_j\ra$
non-zero that places the system away from half-filling. Therefore half-filling corresponds to the
{\em time-reversal symmetry} for the rotor dynamics which pins down the 
Lagrange multiplier $\lambda$ to zero.

\begin{figure}[t]
\includegraphics[width=0.7\linewidth]{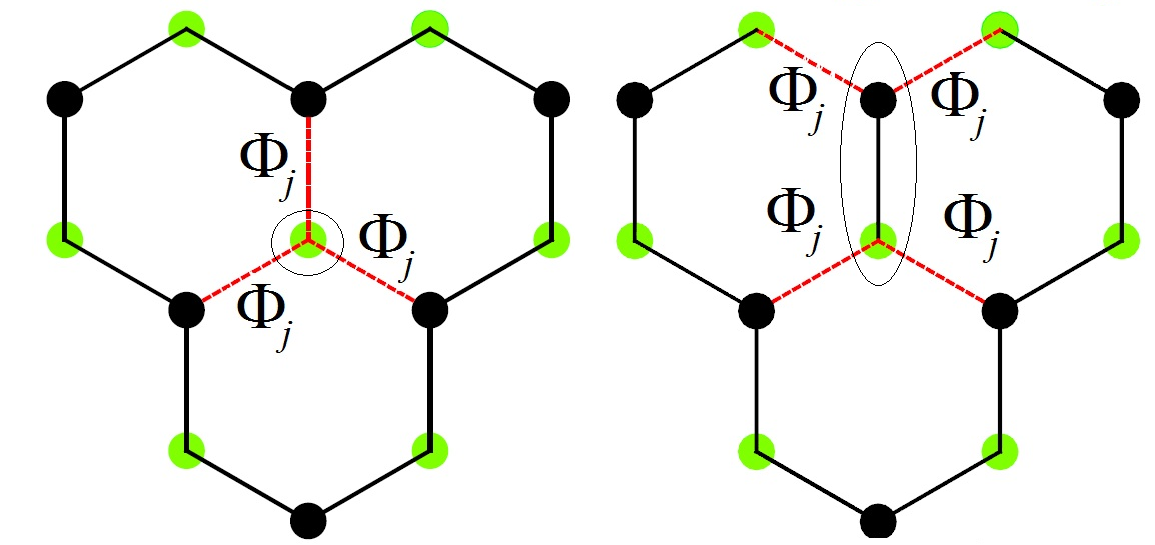}
\caption{(Color online): Two possible clusters for the solution of
   the slave rotor. The dotted lines indicate the mean field value of
   the rotor field connecting the cluster at hand to its neighbours.
}
\label{boththem.fig}
\end{figure}

Now the remaining challenge is to solve the rotor problem posed by
the Hamiltonian~\eqref{rotor.eqn}. One simple way to think of the rotor Hamiltonian
is to fix a value for $\ell$ that gives a local Hilbert space 
dimension of $2\ell+1$ for each rotor.
This Hilbert space grows as $N^{2\ell+1}$ where $N$ is the number of lattice sites.
However if we further decouple the nearest neighbour terms as
\be
   e^{-i\theta_i+i\theta_j}\approx e^{-i\theta_i}\la e^{i\theta_j}\ra +{\rm h.c.}
   =e^{-i\theta_i} \Phi_j +{\rm h.c.},
\ee
where a mean field rotor variable $\Phi_j=\la e^{i\theta_j}\ra $ is introduced,
the rotor problem is considerably simplified. This mean field decomposition can be
implemented on various clusters.
We consider two types of clusters with finite number of sites denoted
in Fig.~\ref{boththem.fig}. For the single-site cluster the mean field
variables connect a given site to its neighbours on the lattice,
and the single-site Hamiltonian has no structure. In this case the
mean field rotor Hamiltonian for every site is given by,
\begin{align}
   H_\theta^{\rm 1-site~MF}=-3\chi_f\Phi \left( e^{-i\theta}+e^{i\theta} \right)
   +\frac{U}{2}\left(n^\theta\right)^2
\end{align}
where the coefficient of $3$ is due to three neighbours of every single-site
(Fig.~\ref{boththem.fig}, left side). The explicit matrix representation of the
above single-site mean field Hamiltonian is,
\be
\left[\begin{array}{ccccc}
   u\ell^2	&-3\chi_f\Phi & 0 &\ldots & 0\\
  -3\chi_f\Phi^* &u(\ell-1)^2 & -3\chi_f\Phi &\ldots &0\\
   0 & -3\chi_f\Phi^* &\ddots &\ldots   &\vdots\\
   \vdots & \vdots & -3\chi_f\Phi^*  &u(-\ell+1)^2 & -3\chi_f\Phi \\
   0 & 0 &\ldots &-3\chi_f\Phi^* & u(-\ell)^2
      \label{matrix.eqn}
\end{array}\right]
\ee
where for notational brevity we have introduced $u=U/2$.
Within the mean field a two-site cluster can also be adopted
(see Fig.~\ref{boththem.fig}) for which the Hamiltonian becomes,
\begin{align}
   &{H}_\theta^{\rm 2-site~MF}=-\chi_f\left( {{e}^{-i{{\theta }_{1}}}}
{{e}^{i{{\theta }_{2}}}}+{{e}^{-i{{\theta }_{2}}}}{{e}^{i{{\theta }_{1}}}}
\right)\notag\\&-4{{\chi }_{f}}\Phi \left( \cos {{\theta }_{1}}+\cos
 {{\theta }_{2}} \right)+u{{\left( n_{1}^{\theta }
 \right)}^{2}}+u{{\left( n_{2}^{\theta } \right)}^{2}}.
 \end{align}
The above Hamiltonian operates in the two-particle space represented by
$|\nt_1,\nt_2\ra$ where $\nt_i=-\ell_i\ldots\ell_i$ for $i=1,2$. The effect
of the two-site MF Hamiltonian on every such state is given by,
\begin{align}
   &-\chi_f\left( |\nt_1-1,\nt_2+1\ra+ |\nt_1-1,\nt_2+1\ra\right)\nn\\
   &-2\chi_f\Phi \sum_{a=\pm1}\left(|\nt_1+a,\nt_2\ra+|\nt_1,\nt_2+a\ra \right)\nn \\
   &+u\left({\nt_1}^2+{\nt_2}^2\right)|\nt_1,\nt_2\ra.
\end{align}
The following algorithm self-consistently determines all the
mean field parameters: For given set of external parameters such as $U$,
(I) Start with an initial guess for $\chi_f$.
(II-a) For the above $\chi_f$, guess a $\Phi$. (II-b) Diagonalize the matrix~\eqref{matrix.eqn}
and obtain its ground state. (II-c) In the obtained ground state 
update the $\Phi=\la e^{i\theta}\ra$
and keep repeating until $\Phi$ is self-consistently determined.
(III) For the present value of $\chi_f$ and $\Phi$, use Eq.~\eqref{chitheta.eqn} to obtain $\chi_\theta$.
(IV) Plug in the $\chi_\theta$ into the spinon Hamiltonian~\eqref{spinon.eqn} and diagonalize it.
(V) For the ground state of the above spinon Hamiltonian use Eq.~\eqref{chif.eqn} to update
the initial guess $\chi_f$. This procedure is repeated until mean field parameters
$\chi_f,\chi_\theta,\Phi$ are self-consistently determined.
It must be noted that the above procedure is done for a fixed value of $\ell$. One has to
repeat the procedure for larger values of $\ell$ to ensure that the final converged 
results do not change much upon further increase in the dimension of the rotor space. 
We confirm that as previously noted~\cite{Paramekanti} the choice $\ell=2$ is accurate enough.

Upon increasing the Hubbard interaction $U$ beyond a critical point $U_b$ the mean field
parameter $\Phi$ vanishes that corresponds to strong fluctuations in the phase
variable, and hence frozen fluctuations of the corresponding number operators,
i.e. {\em putative} Mott state. However, as will be discussed in the next section
the phase fluctuations have still have the chance to survive in the form of 
{\em sign fluctuations} that can be captured by enslaving an Ising variable.

\section{slave spin method}
The charge degree of freedom at every site can be described by variety
of methods. Description in terms of a rotor variable whose conjugate variable
controls the charge state is one possibility. Another appealing possibility is
to use an Ising variable to denote the charge state. Using an Ising variable
to specify the electric charge can be implemented in various ways~\cite{Sigrist,HasanSlaveSpin,Si}.
In this section we briefly review the presentation of Ref.~\cite{Sigrist} and adopt
it in our investigation of the ionic Hubbard model on the honeycomb lattice.

In this representation an Ising variable $\tau$ is introduced to take care of
the charge configuration of every site. Empty and doubly occupied configuration
are represented by $|+\ra$, while singly occupied configurations possessing 
local moment both are represented by $|-\ra$ where,
\be
   \tau^z |\pm\ra = \left(\pm 1\right) |\pm\ra
\ee
The correspondence between physical states and those in the Hilbert space
extended by introduction of Ising variables is:
\begin{align}
   & \left| {\rm empty} \right\rangle =\left| + \right\rangle \left| 0 \right\rangle
  \nonumber\\
 & \left| {\rm singly~occupied},\up \right\rangle =\left| - \right\rangle \left| \up  \right\rangle  \nonumber\\
 & \left| {\rm singly~occupied},\down \right\rangle =\left| - \right\rangle \left| \down  \right\rangle  \nonumber\\
 & \left| {\rm doubly~occupied} \right\rangle =\left| + \right\rangle \left| \up\down \right\rangle.
  \label{IsingStates.eqn}
\end{align}
where states on the left hand correspond to physical electrons, and those in the
right hand are product of states corresponding to Ising variables, $|\pm\ra$,
and those corresponding to spinon. Creation of each electron has almost a parallel on the
right side implying that $c^\dagger\sim f^\dagger$. However each time an electron is
created, the charge state flips between the one having a local moment and the one having no
local moment. This corresponds to a flip in the Ising variable that can be achieved with
the action of Pauli matrix $\tau^x$. Therefore the physical electron at every site $j$
can be represented as,
\begin{align}
   c_{j\sigma }^{\dagger }\equiv \tau^x_j f_{j\sigma }^{\dagger }.
   \label{slaveIsing.eqn}
\end{align}
where $f^\dagger_{j\sigma}$ creates a spinon of spin $\sigma$ at site $j$.
The fact that creation of each physical charge is synonymous to creation of
a spinon implies that the physical charge is basically carried by spinons.
This was a key observation made by Nandkishore and coworkers~\cite{Nandkishore} that
is formally reflected in Eq.~\eqref{slaveIsing.eqn} as the fact that the Pauli
matrix $\tau^x$ being a real matrix can not absorb the $U(1)$ phase transformation that generates 
the conservation of charge, and hence all the charge of electron is carried by the spinon.

The Hilbert space represented by the product of Ising pseudo-spin and spinon spaces
is larger than the physical space and includes states such as $|+\ra|\up\ra$, etc that do not 
correspond to any physical state. Inspection shows that those states can be
eliminated by the following constraint:
\be
   \tau^z_j+1-2(n_j-1)^2=0.
   \label{Isinggheid.eqn}
\ee
In this new representation the ionic Hubbard Hamiltonian becomes,
\begin{align}
   &H=-\sum\limits_{\la ij \ra}\tau_i^x \tau_j^x f_{i\sigma}^{\dagger} f_{j\sigma}
+\frac{U}{4}\sum_j{\left( \tau _{j}^{z}+1 \right)}\nn\\
&+\Delta \sum_j f^\dagger_{j\sigma} f_{j\sigma} (-1)^j
\end{align}
where as before the hopping $t$ of the physical electrons is set as the unit of
energy and the constraint Eq.~\eqref{Isinggheid.eqn} has enabled us to cast the Hubbard $U$
into a form involving only Ising variable $\tau^z$.
Again in the ionic potential term the Ising pseudo-spins being squared to unit matrix
cancel each other and hence the staggered ionic potential is only coupled to the
spinons. Mean-field decoupling of the Ising and spinon variables
is lead to two separate Hamiltonians governing the dynamics of spinons $f$ and
Ising variables $\tau$ as follows:
\begin{align}
   H_f=&-\sum\limits_{ij}{{{\chi }_{I}}f_{i\sigma }^{\dagger }{{f}_{j\sigma }}}
+\Delta \sum_j f^\dagger_{j\sigma} f_{j\sigma} (-1)^j\nn\\
   &-2\lambda'\sum_j\left(f^\dagger_{j\up}f_{j\up}+f^\dagger_{j\down}f_{j\down}-1\right)^2,
\label{HfIsing.eqn}
\end{align}
and
\begin{align}
   H_{\rm ITF}=-\chi'_f\sum\limits_{ij}\tau _{i}^{x}\tau _{j}^{x}
   +\left(\frac{U}{4}+\lambda'\right)\sum\limits_{i}{\tau _{i}^{z}}
   \label{Htau.eqn},
\end{align}
where the two Hamiltonians are coupled to each other through the following
self-consistency equations:
\bearr
   &&\chi_I= \la \tau^x_i \tau^x_j\ra,\label{chiI.eqn}\\
   &&\chi'_f=\la f^\dagger_{i\sigma} f_{j\sigma}\ra,
\eearr
and the Lagrange multiplier $\lambda'$ is introduced to implement the
constraint~\eqref{Isinggheid.eqn} on average. As can be seen, within
the present representation, the fermion part,~\eqref{HfIsing.eqn}
still remains an interacting problem of the Hubbard type, where the scale of on-site interaction
among the spinons is set by the Lagrange multiplier $\lambda'$.  The mean field
approximation lends itself on the assumption that the system in the enlarged
Hilbert space is a product state composed of a spinon part and an Ising part.
The spinon part of the wave function is expected have a form close to an Slater
determinant and hence the parameter $\lambda'$ is expected to represent a 
small residual interactions between the spinons. Approximate strategies to
handle the interacting spinons have been suggested and discussed in Ref.~\cite{Sigrist}.
With this argument we reckon that it is reasonable to assume $\lambda'=0$ as an
approximate strategy to obtain the simplest possible solution~\cite{Fiet}.

To understand the nature of the approximation $\lambda'\approx 0$, let us 
focus on some special lucky situations where it can be proven that $\lambda'=0$ is
exact. When the ionic term is absent, i.e. for the pure Hubbard model one may 
have situations where the partition function $Z$
happens to be an even function of $U$, such as the particle-hole symmetric
case considered here~\cite{michele2011,michele2015}.
In this case, first of all the Hubbard interaction at half-filling can be
written as,
\be
   Un_{j\up} n_{j\down} -\frac{U}{2}\left(n_{j\up}+n_{j\down} \right),
   \label{phsh.eqn}
\ee
where due to half-filling condition the chemical potential $\mu=U/2$ is used
and the site index has been dropped for simplicity.
Under a particle-hole transformation in one spin sector only
(let us call it PH$\sigma$ transformation in this paper), namely,
\be
c^\dagger_{j\up}=\tilde c^\dagger_{j\up},~~~c^\dagger_{j\down}=(-1)^j \tilde c_{j\down},
\ee
where $(-1)^j=\pm 1$ depends on whether it is on sublattice A, or B. 
Under PH$\sigma$ the role of charge
and spin density are exchanged, namely $n_\up+n_\down \to \tilde n_\up-\tilde n_\down$.
This transformation maps the particle-hole symmetric Hubbard interaction of
Eq.~\eqref{phsh.eqn} to,
\be
   \tilde U\tilde n_{j\up} \tilde n_{j\down} -
   \frac{\tilde U}{2}\left(\tilde n_{j\up}+\tilde n_{j\down} \right)
\ee
which is nothing but the original Hubbard interaction at half-filling with the
only difference that $\tilde U=-U$. At half-filling under the PH$\sigma$ transformation
is a symmetry of the Hubbard model which implies properties of system are even with respect
to Hubbard $U$.

\begin{figure}[t]
\includegraphics[width=0.95\linewidth]{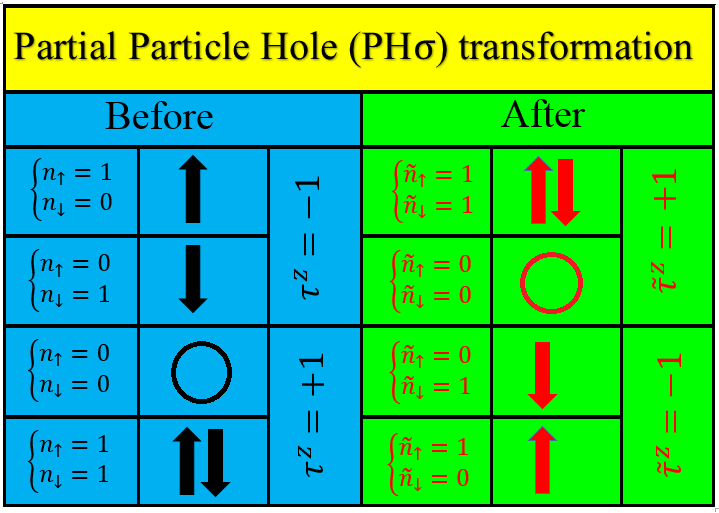}
\caption{(Color online): Schematic summary of partial particle-hole
transformation, PH$\sigma$ that affects only down-electrons.
It basically exchanges the local charge and spin densities.
In a setting with one Anderson impurity, the PH$\sigma$ would correspond to
transformation between spin and charge Kondo effects.
}
\label{lagrange.fig}
\end{figure}

Let us now examine how does the constraint~\eqref{Isinggheid.eqn} behave under
the PH$\sigma$ transformation. As indicated in the Fig.~\ref{lagrange.fig}
the role of PH$\sigma$ transformation in terms of Ising pseudo-spins is to
exchange the role of $|+\ra$ and $|-\ra$ states of Ising variables.
Therefore the effect of PH$\sigma$ transformation on any operator ${\cal O}$
that contains Ising variables is to change ${\cal O}\to \tau^x {\cal O}\tau^x$.
This transformation leaves the first term in the ITF Hamiltonian~\eqref{Htau.eqn} intact,
but it changes the second term at a given site as follows:
\bearr
   && \left(\frac{U}{4}+\lambda'\right) \tau^z \to
   \left(\frac{U}{4}+\lambda'\right) \tau^x\tau^z\tau^x\nn\\
   &&=  -\left(\frac{U}{4}+\lambda'\right)\tau^z
   =-\left(-\frac{U}{4}+\lambda'\right)\tau^z,
\eearr
where in the last equality we have used the fact that the Hubbard model at half-filling is
even with respect to Hubbard $U$. Therefore we have proven that {\em at half-filling for the Hubbard model,
the Lagrange multiplier $\lambda'$ is exactly zero}.

Turning on the ionic potential $(-)^j \Delta (n_\up+n_\down)$, the PH$\sigma$ transformation
maps to $(-)^j\Delta (\tilde n_{j\up}-\tilde n_{j\down})$, i.e. the charge density is
mapped to spin-density. Therefore the PH$\sigma$ is not a symmetry of {\em ionic} Hubbard model
at half-filling. Hence the partition function has both even and odd parts as a function of $U$.
This prevents the Lagrange multiplier $\lambda'$ from becoming zero. However the above symmetry
consideration suggests that the physics of ionic (and repulsive) Hubbard model maps
onto the physics of attractive Hubbard model in the presence of a staggered magnetization
field (since it is coupled to spin density in staggered way). Another merit of the above
symmetry discussion is that based on the following argument in Ref.~\cite{michele2011},
we can infer what is precisely missed by {\em the approximation} $\lambda'=0$:
Let us rewrite the constraint~\eqref{Isinggheid.eqn} as,
\be
   {\cal P}_+=1+\tau^z_j\Omega_j=0,~~~\Omega_i=1-2(n_j-1)^2,
\ee
which identifies the operator $\Omega_j$ as the fluctuations of
the charge away from half-filling. It was shown in Ref.~\cite{michele2011} that
to all orders in perturbation theory, the term $1$ in ${\cal P}_+$ contributes
only in even powers of $U$ while the second term contributes only in odd powers of $U$.
Therefore for a half-filled situation of the pure Hubbard model, the effect of second term
is nullified, and basically one need not worry about projection. This is another
way of saying that the Lagrange multiplier $\lambda'$ at half-filling becomes zero.
By adding ionic term to the half-filled Hubbard model, or placing the Hubbard model
itself away from half-filling, $\lambda'$ is expected to be small based on our earlier
argument on small residual interactions. In this case using the approximation 
$\lambda'\approx 0$ amounts to missing the effects that are odd functions of the Hubbard $U$.

\begin{figure}[t]
\includegraphics[width=0.4\linewidth]{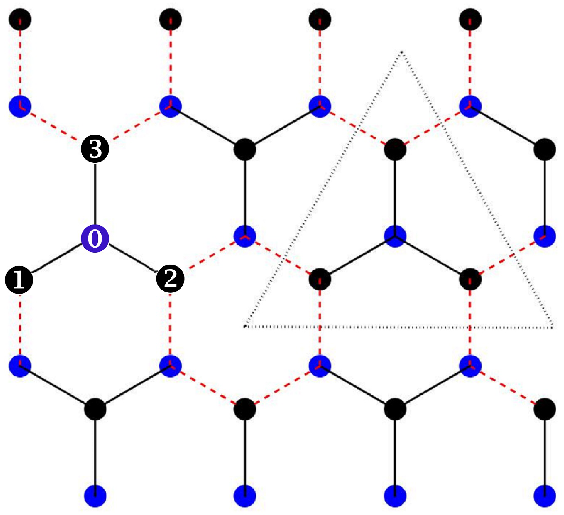}
\includegraphics[width=0.4\linewidth]{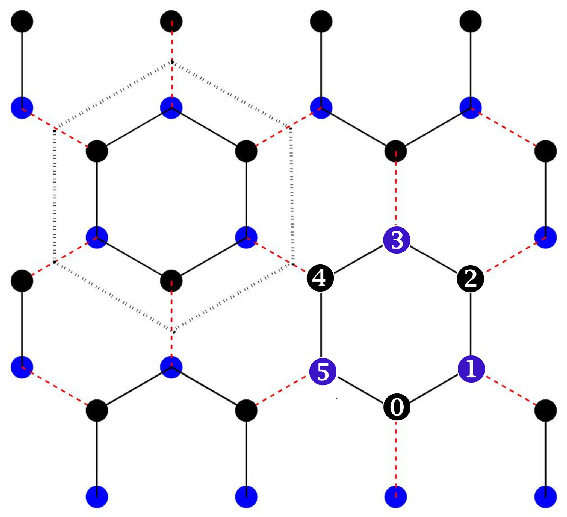}
\caption{(Color online) Two choices for the cluster mean field
treatment of the slave Ising pseudo-spins; 
the Y shaped (left) and hexagonal cluster (right).}\label{zx}
\end{figure}

Within the approximation of $\lambda'=0$, the spinon part describes a non-interacting band of
spinons whose bandwidth is renormalized by the $\chi_I$ parameter obtained from
the Ising part. The physics of transition to orthogonal state is then captured by the disordering
transition of the Ising sector~\eqref{Htau.eqn} where vanishing quasi-particle
weight of the physical electrons are characterized by $\la \tau^x\ra =0$~\cite{Nandkishore,michele2015}.

In the absence of the ionic term, i.e. when $\Delta=0$, both ordered and disordered
side of the ITF Hamiltonian~\eqref{Htau.eqn} are conducting state: (i) If the underlying
rotors are ordered, namely $\la e^{i\theta}\ra \ne 0$ the conducting state within the
Hubbard model is a Fermi liquid. This holds for $U<U_b$. (ii) When the rotor order vanishes
beyond $U_b$, i.e. $\la e^{i\theta}\ra =0$ still we may have $\la \tau^x\ra\ne 0$ which 
again is a conducting state (Ising metal) for the pure Hubbard model. This state
persists until $U_\perp$ where the Ising order vanishes and the pure Hubbard model
describes the orthogonal metallic state. By adding the ionic term $\Delta$, it
is crucial to note that on-site ionic potential does not couple to neither slave Ising
variables, nor to the salve rotor variables. However, the effect of the ionic potential $\Delta$
is to modify the order parameters through the mean field self-consistency equations, but
their order-disorder physics remains the same as the Hubbard model as the
ionic term does not explicitly appear in the Ising or rotor sectors. The essential role
of the ionic term is to create Bragg reflections in the spinon Hamiltonian and to gap 
them out which corresponds to rendering Fermi liquid, Ising metal and orthogonal metal
phase of the pure Hubbard model to band insulator, Ising band insulator, and orthogonal
insulator, respectively. 

The disordering phase transition of the Ising Hamiltonian~\eqref{Htau.eqn} can be captured
within a simple cluster mean field approximation. Let us decompose the lattice to clusters 
$\Gamma$ labeled by integer $I$ whose internal sites are labeled by integers $a,b$ etc.
Then the Ising variables are denoted by $\vec\tau_{\Gamma,a}$. With this re-arrangements,
and after mean field decoupling of the cluster with its surrounding sites via a mean field
order parameter $m=\la \tau^x\ra$ (see Fig.~\ref{zx}), the Ising part becomes,
\be
   H=\sum_\Gamma H_\Gamma,
\ee
where the Hamiltonian for cluster $\Gamma$ is,
\begin{align}
   H_\Gamma= -\chi'_f\sum_{a,b\in \Gamma} \tau_{\Gamma,a}^x\tau_{\Gamma,b}^x
   -\frac{mz}{2}\sum_{a\in\Gamma}\tau_{\Gamma,a}^x+\frac{U}{4}\sum_a{\tau _{\Gamma,a}^{z}}.
   \label{Hcluster.eqn}
\end{align}
In this cluster Hamiltonian, the $z$ denotes number of bonds crossing the
boundary of the cluster. The factor $1/2$ avoids double counting and $m=\la\tau^x\ra$ is the
Ising order parameter that at the mean field level decouples the cluster $\Gamma$ from its
surroundings, but the interactions within the cluster $\Gamma$ are treated with exact diagonalization.
In Fig.~\ref{zx} we have depicted clusters used in the present work. For more
details on the construction of the Hilbert space and diagonalization of the Hamiltonian,
please refer to the Appendix. The Ising disordering considered here for 4-site and 6-site clusters
do now show appreciable difference. In this work we report the critical values 
$U_c$ of the Ising disordering transition that is obtained from 6-site clusters.

\section{Results}

\begin{figure}[t]
\includegraphics[width=0.47\linewidth]{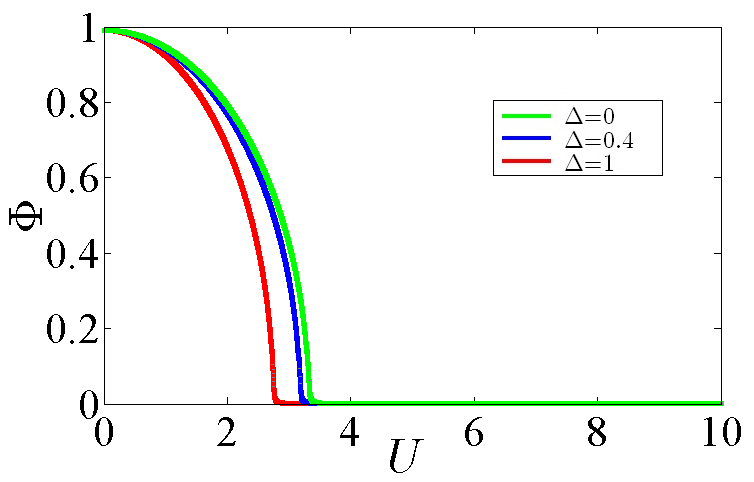}
\includegraphics[width=0.52\linewidth]{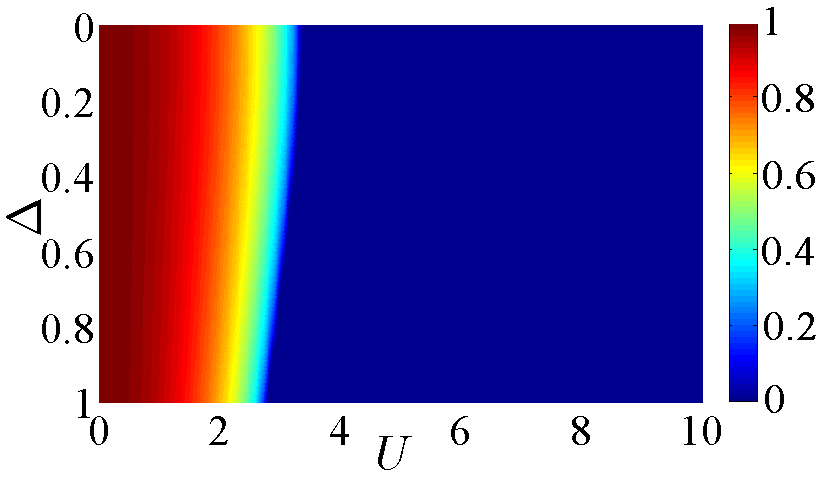}
\caption{(Color online) The slave rotor order parameter $\Phi$. The right
   panel shows the intensity plot in the plane of Hubbard $U$ and ionic
   potential $\Delta$ on the honeycomb lattice. The left panel shows the
   dependence of $\Phi(U)$ for some selected values of $\Delta$ as indicated
   in the legend. The rotor (single boson) condensate is lost at $U_b$.
}
\label{rotor.fig}
\end{figure}

In Fig.~\ref{rotor.fig} we have plotted the results of a slave-rotor mean field
for a two-site cluster. The rotor Hilbert space in this plot has been constructed
for the angular momentum $\ell=2$. We have checked that the results are not sensitive
to increase in the size of the rotor Hilbert space beyond $\ell=2$.
The left panel shows the evolution of order parameter
$\Phi$ of rotors as a function of Hubbard $U$ for a selected set of ionic potentials
indicated in the legend. As can be seen for $\Delta=0$ the critical value $U_b$
starts around $3.5$ and decreases by increasing $\Delta$. At $\Delta=1$ the critical
value for the disordering of single-boson is around $2.8$ . This means that in the
presence of a staggered potential it becomes easier to loos the single-boson
condensate whereby the Ising phase (paired boson superfluid) starts. 
Right panel in the figure
provides an intensity map of the rotor order parameter in the $(U,\Delta)$ plane.
The blue region corresponds to zero single-boson condensation amplitude, and the
red corresponds to maximal (i.e. 1) condensation amplitude for the single-bosons
$\la e^{i\theta}\ra$.

\begin{figure}[t]
\includegraphics[width=0.47\linewidth]{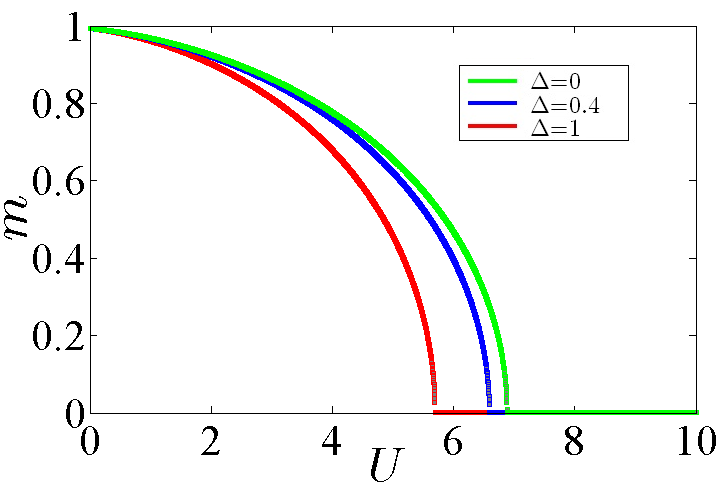}
\includegraphics[width=0.52\linewidth]{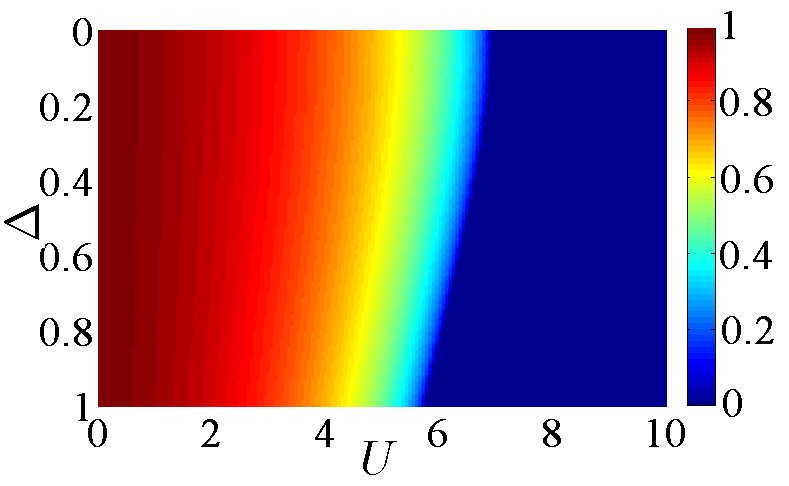}
\caption{(Color online) The slave spin order parameter $m=\la \tau^x\ra$.
The right panel shows the intensity plot of the Ising order parameter $m$ as a function of
$U$ and $\Delta$. Right panel shows the variations in the Ising order parameter as a
function of $U$ for some selected values of $\Delta$ indicated in the legend.
The data are obtained within the cluster mean field approximation for a 6-site cluster.
}
\label{ising.fig}
\end{figure}

In Fig.~\ref{ising.fig} we present the cluster mean field results for the
Ising order parameter. The left panel shows the Ising magnetization $m=\la \tau^x\ra$
as a function of $U$ for selected values of the staggered potential $\Delta$ indicated
in the figure.
The right panel provides an intensity map of the Ising order parameter
in the $(U,\Delta)$ plane.
The color code is the same as in Fig.~\ref{rotor.fig}.
Once the Ising order goes away, we are in the orthogonal phase.
By increasing the staggered ionic potential from
$\Delta=0$ to $\Delta=1$, the critical value $U_c$ decreases from $\sim 6.9$
to $5.7$. This trend is similar to the behavior of rotor order parameter,
i.e. the effort of $U$ to destroy the Ising order parameter $m=\la \tau^x\ra$ is
assisted by the ionic potential $\Delta$. This points to the fact that 
getting both IBI and OBI is facilitated by ionic potential $\Delta$. 
Larger the ionic potential, easier to "quantum melt" the 1- and 2- boson condensates
that correspond to entering Ising and orthogonal phases.

\begin{figure}[t]
\includegraphics[width=0.7\linewidth]{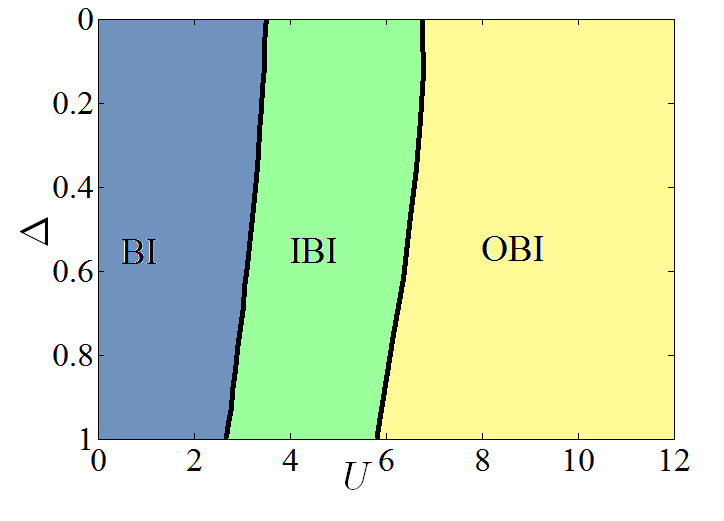}\\
\caption{(Color online): Phase diagram of the ionic Hubbard model within the
present combination of slave-rotor and slave-spin mean field approximations.
}
\label{ibi-obi.fig}
\end{figure}

Fig.~\ref{ibi-obi.fig} combines Figs.~\ref{rotor.fig} and~\ref{ising.fig} and
shows that there is a clear region $U_b \le U < U_\perp$ where the
rotor order parameter is zero, i.e. the single boson condensate has vanished,
while the Ising order parameter is non-zero, i.e. the double boson is
condensed~\cite{Nandkishore}. In the region $U<U_b$ the underlying
metallic state is a normal fermi liquid which is 
gapped out by directly coupling to the staggered potential and therefore
the underlying Fermi liquid state becomes a normal band insulator.
For $U_b<U<U_\perp$ the 1-boson condensate vanishes, and interaction
between the bosons leads to pairing of bosons and the 2-boson condensates
forms. In this region due to formation of two-boson pairs that admit a $Z_2$ gauge structure, 
the charge fluctuations are controlled by Ising variable which is ordered and gives
IBI. The Ising order parameter vanishes at $U_\perp$ beyond which the
semiconducting transport will be entirely done by spinons. 
For largest values of $U>U_{\rm Mott}$ the system eventually becomes Mott
insulating~\cite{Ebrahimkhas}.

The existence of a
region where the Ising variable is ordered, but the rotor variable is disordered
endows the non-Mott phase of the ionic Hubbard model with a condensate of 
paired charge bosons whence charge fluctuations survive in the form of Ising variables. 
In the ionic Hubbrard model this corresponds to Ising band insulating phase where
the kinetic energy of spinon Hamiltonian is controlled by an Ising order parameter,
and hence the band properties of such a semiconducting phase inherits characteristic
temperature and field dependence from the underlying Ising model. 
Across the Ising transition, the quasi-particle weight of the physical electron is lost,
and the electric charge is carried by spinons which corresponds to loosing the Ising condensate. 
This is the orthogonal phase which in the ionic Hubbard model corresponds to OBI. 
The quasiparticle weight of the physical electrons in this phase vanishes as it is
controlled by the Ising order parameter~\cite{Nandkishore}, and hence in the OBI
phase the states at the bottom of the conduction and top of the valence band 
are not visible by ARPES. 
However, since the current operators is solely
constructed by the spinons, the optical conductivity (i.e. the current-current correlation
function) does couple to the states near the bottom of conduction band and those near the 
top of the valence band of the resulting OBI.
Therefore an important characteristic property of OBI is that the optical conductivity
gap is expected to be smaller than the ARPES gap. For the two-dimensional
semiconductors or insulators the ability to tune the chemical potential into the
conduction band provides a chance to examine such an "ARPES-dark stats" by quantum oscillations
experiments. Once the chemical
potential is tuned to the conduction band, the thermally excited carriers into
the ARPES-invisible states at the bottom of the conduction band would display
quantum oscillations. The "ARPES-dark" states of the OBI phase would couple
to thermal probes as well which means that the gap extracted from thermal
measurements will be smaller than the ARPES gap. 

\section{Discussions and summary}
We have investigated the phase transitions of the ionic Hubbard model on the
honeycomb lattice using a combination of slave rotor and slave spin mean field
theories. The phase diagram of the ionic Hubbard model on the honeycomb lattice
within the present method is shown in Fig.~\ref{ibi-obi.fig}.
For small values of Hubbard $U<U_b$ (blue phase) we find a normal band insulating
(semiconducting) state. For intermediate values of $U_b<U<U_\perp$ (green phase) the
properties of the band insulating phase is controlled by an Ising condesate amplitude.
In this band insulator the rotor is disordered, $\la e^{i\theta}\ra=0$,
but the Ising variables remains ordered, $\la \tau^x\ra\ne 0$. 
The interesting nature of this phase lends itself on the chargon pairing~\cite{Nandkishore}.
By further increasing $U$ the Ising order is lost, and we end up in even more
exotic orthogonal phase (yellow phase) where the chargon pair condensate has
vanished and hence semiconducting transport is dominated by spinons. 
The spinons inherit the electric charge of the electron. This is simply because
the $U(1)$ symmetry (charge conservation) of the original ionic Hubbard model can not be incorporated
into the Ising pseudo-spins (as they are real matrices), 
the $f$ operator inherits the electric charge of electrons and 
a set of conduction-valence bands of spinons is left for semiconductor transport.
Therefore yellow phase in Fig.~\ref{ibi-obi.fig} can be viewed as a {\em spinon semiconductor}. 

Now suppose that we are given a semiconducting (band insulating) sample. How do we differentiate
whether it is BI, IBI, or OBI?
(i) Let us start with OBI (yellow region in Fig.~\ref{ibi-obi.fig}): 
Since vanishing of the Ising order
amounts to loosing the quasi-particle weight of the physical electron, the defining property 
of the orthogonal (yellow) phase is that
the ARPES gap is larger than the optical gap as the former probe couples to electrons
whose quasiparticle weight is lost in the orthogonal phase while the later probe
couples to the spinon current operator. The same holds for the thermal gap.
In the semiconducting phase the gap can be extracted from thermal measurements as well.
Again this is expected to be smaller than the ARPES gap which signals existence of 
ARPES-dark states which are nothing but the spinon states. This criterion not only
qualifies a given sample as OBI, but also from fundamental physics point of view can
serve as a proof of quantum number fractionization phenomena.

(ii) The essential property of the Ising phase (green region in Fig.~\ref{ibi-obi.fig}) 
is that it depends on an Ising order parameter $m=\la\tau^x\ra$.
Due to the temperature dependence of underlying (slave) Ising order parameter
that multiplies the kinetic energy of spinons, the effective mass will
correspondingly acquire a temperature dependence characteristic of the 
Ising order. This can be detected by standard cyclotron resonance experiments
and monitoring their temperature dependence $m^*(T)$. 
Particularly when the temperature is high enough to hit the "thermal"
disordering point of the Ising variable $\la \tau^x\ra$, or when $U$ is close 
to  $U_\perp$, the cyclotron effective
mass is expected to be enhances as one approaches the OBI phase from the IBI side.
The characteristic Ising power-laws of the Ising universality class are expected
to leave their footprint in the temperature dependence of the effective mass.
This situation is in sharp contrast to normal semiconductors where band parameters are almost
rigid and do not depend on the temperature. 
In normal semiconductors the dominant temperature
dependence determining the transport properties appears in the density of excited carriers,
while in the Ising semiconductor, in addition to the carrier density,
every property involving the Ising order parameter acquires an additional and distinctive temperature
dependence. This can serve not only to distinguish orthogonal semiconductors/insulators
from their normal relatives, but also as a existence proof for the underlying Ising variable and hence
the fractional nature in two-dimensional semiconductors with strong correlations.

(iii) The Ising phase may have anomalous response to applied magnetic fields.
Although the Ising variable $\tau^z$ labels the charge states, nevertheless
it carries information about the local moments. The $\tau^z=+1$ state
carries no net magnetic moment and hence in the first order, it does not couple
to an external magnetic field $B$. However the $\tau^z=-1$ charge state carries a 
net local moment and hence can Zeeman couple to $B$ to gain energy. In this
way, the applied $B$ field effectively couples to Ising condensate. This may provide
extra sensitivity to $B$ field in Ising semiconductors as opposed to normal
semiconductors. Given that the resistivity of rare earth monochalcogenides is very 
sensitive to applied magnetic fields, and that the heavy Fermion elements involved
are qualified for strong correlations, we suspect that rare earth semiconducting 
systems such as Europium monochalcogenides EuX or Samarium monochalcogenides SmX 
where X=S, Se, Te~\cite{Smirnov} and rare earth nitrides~\cite{Sclar} can be 
interesting platforms to search for signatures of underlying Ising condensate.

\begin{figure}[t]
\includegraphics[width=0.7\linewidth]{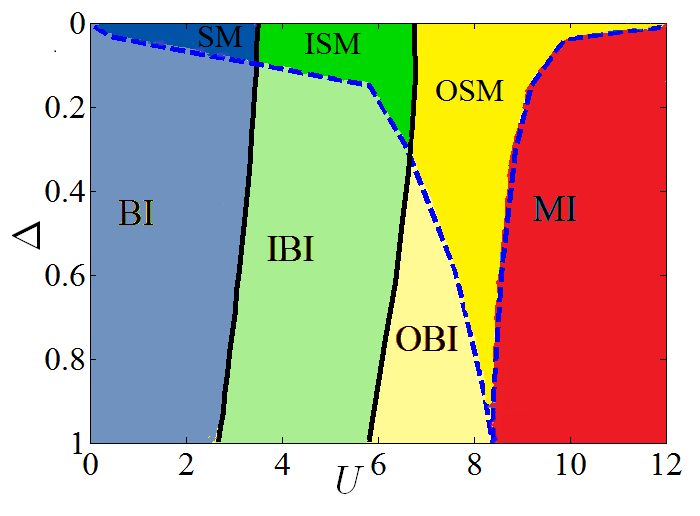}
\caption{(Color online): 
Comparison between the present mean field phase diagram and our previous
DMFT phase diagram. The dashed lines correspond
to DMFT results. Competition between $U$ and $\Delta$ in DMFT scenario gives
a massless Dirac phase between the two dashed lines. Rotor and Ising 
transitions partition the massless Dirac phase into semi-metal (SM), Ising
semi metal (ISM) and orthogonal (SM). 
For discussions see the text. 
}
\label{orthogonal.fig}
\end{figure}

Let us briefly discuss the connection of the present work to other works
on the ionic Hubbard model. Investigations of the nature of intermediate
phase in the ionic Hubbard model fall into two major groups: First group 
suggests that the intermediate phase is gapped, while the second group
suggests gapless intermediate state. The present work also does find a gapped intermediate
phase. However, the gap in the present case is due to symmetry breaking in a 
fractional degree of freedom. This order does not correspond to any spin or
charge density as no form of density operator depends on the Ising pseudo-spin simply 
because $\tau^x$ squares to unit matrix. The gap in the Ising and orthogonal semiconductor
is caused by the Bragg reflection of spinons, and as such there are no
low-energy Goldstone modes associated with our present proposal. 
The second group of investigations suggest a gapless state for a region $\Delta\sim U$. 
Within the present mean field approach, we get three gapped phases depicted 
in Fig.~\ref{ibi-obi.fig}: BI (blue), IBI (green) and OBI (yellow. 
The present approach does not capture the Mott phase as in the mean field
and within the half-filled Hubbard model we do not take interactions among
spinons of the IBI phase into account. However a comparison between our
previous dynamical mean field theory (DMFT) result is rewarding~\cite{Ebrahimkhas}.
Within the DMFT approach the battle between $U$ and $\Delta$ to close the gap takes place.
The dashed lines in Fig.~\ref{orthogonal.fig} represent the phase boundaries
from DFMT. The left branch of the dashed line separates band insulator from
semi metal (SM) while the right branch of the dashed phase boundary separates SM
from Mott insulator (MI). The intermediate phase is a massless Dirac phase within
the DMFT. When we superimpose the DMFT phase diagram~\cite{Ebrahimkhas} with that of
Fig.~\ref{ibi-obi.fig} we find that the phase boundaries obtained from present
study (bold lines) partition the BI and SM phase of the DMFT phase diagram into
three phase corresponding to normal, Ising and orthogonal variants. Although
these are two different methods, and critical values obtained from DMFT
and present studies maybe correspond to different mechanisms, but that does
not concern us here. Improvements in the approximations may push the bold lines
slightly away, but does not change the fact that the bold phase boundaries
cross the left branch of the DMFT (dashed boundary). This comparison sheds a new
light: Realization of IBI and OBI phases requires large enough ionic potential.
For very small ionic potentials, the DMFT battle between $U$ and $\Delta$ can possibly  
kill the insulating phase, and give a massless SM. Then increase in Hubbard $U$
will give rise to Ising semi metal (ISM) or orthogonal semi metal (OSM)~\cite{Chinese}.
If the ionic potential grows further, the green phase in the band insulating side
also gets a chance and therefore IBI could be realized if $\Delta$ is larger than about $\sim 0.1$
(in units of hopping $t$ of course). If we keep increasing $\Delta$ beyond $\sim 0.38$ the
OBI phase also gets a chance. However if the DMFT scenario of battle between $U$ and $\Delta$
suggests that the OBI phase does not directly transform into Mott phase, but instead goes through
an orthogonal semi metal which is appealing: The battle will continue in the fractionalized
OBI phase of spinons and can presumably close the spinon gap in OBI to render it OSM before 
getting into Mott phase~\cite{Ebrahimkhas}. 

Within the present mean field approximate treatment of the IBI-OBI phase
transition, the thermal probes are coupled to 
spinons that are independent of Ising pseudo-spins. The optical probe
on the other hand always couples to the spinons. However
going beyond the mean field by properly taking the fluctuations of internal
gauge fields that glue the spinons to Ising field into account 
is expected to provide corrections to the present picture.
Therefore thorough investigation of gauge fluctuations
and its effect on the physical properties of Ising and orthogonal phases remains and is worth to be explored.
Thinking along the schematic table of Fig.~\ref{schematic.fig} one may also wonder
about other possible columns to start with at $U=0$. An interesting possibility
can be the Ising and orthogonal cousin of the Anderson insulator, where the insulating behavior at $U=0$ is
due to randomness. This will add another interesting aspect to the Mott-Anderson
problem, namely the interplay between the Hubbard $U$ and randomness around $U_c^\perp$,
and possible glassy phases of spinons. This problem is currently under investigation
in our group~\cite{Amini}.

Let us emphasize that although in the present paper we are confined
to zero temperature where
the quantum phase transition between IBI and OBI is driven by Hubbard $U$ by
destroying the Ising condensate; the quantum fluctuations
are not the only way to destroy a condensate. Thermal fluctuations can be
conveniently employed to achieve this goal. Within this scenario, once a system is found
in IBI phase, simply rising the temperature gives a chance to the OBI phase.
If the anomalous magnetic field dependence of the transport properties in 
monochalcogenides~\cite{Sclar,Smirnov} is due to the Ising order, then searching for
"ARPES-dark" states in elevated temperature can support this assumption. 
By increasing the temperature, once the underlying Ising order is lost, 
the ARPES gaps starts to deviate from thermal gap. Moreover since the 
effective mass of spinons in the IBI is controlled by Ising order parameter
the cyclotron mass will acquire a characteristic Ising-like temperature
dependence.

To conclude, {\em additional temperature and magnetic field dependence due to an underlying 
Ising order parameter that is attached to spinons is the key feature of transition to orthogonal phase}. 
This observation suggests that the correlated semiconductors maybe an alternative and 
appealing (if not superior) rout to search for correlation driven phenomena 
where the sensitivity of semiconducting
carrier density to temperature combined with the temperature dependence of underlying
Ising order field cooperate to reveal information about fractional excitation
of solids. Indeed in the absence of tunability of the correlation parameter $U$
in solids, further dependence of the underlying Ising field to temperature
and magnetic field can serve as conveniently tunable parameters to probe 
fractional excitations in correlated semiconductors.
Investigation of inhomogeneity and impurities in spinon-semiconductors 
and their contrast to normal semiconductors can shed light on exotic properties
of spinon semiconductors. From technological point of view, 
given the very extensive use of semiconductors in every day life, 
further exploration of the exotic properties of spinon semiconductors 
may prove useful. 

\section{Acknowledgements}
We thank A G Moghaddam for useful comments and R. Ghadimi for very helpful discussions.
TF appreciates the Ministry of Science, Research and Technology (MSRT) of Iran
for financial support during a visit to Sharif University of Technology.

\appendix
\section{Details of exact diagonalization for clusters}
In this appendix we present details of the exact diagonalization for the ITF 
Hamiltonian on a finite cluster for 4 and 6 site clusters.
We employ group theory methods to reduce the dimension of ensuing matrices.

\subsection{Y-shaped 4-site cluster}
\begin{figure}[t]
\includegraphics[width=0.35\linewidth]{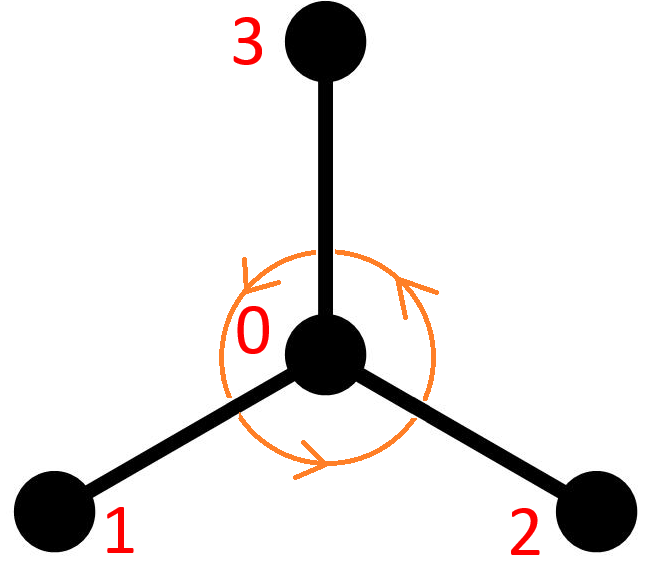}
\caption{The Y-shaped 4-site cluster chosen in the exact diagonalization of
ITF Hamiltonian. The total Hilbert space of this cluster is $2^4=16$ dimensional
labeled by four spins $|\sigma_0,\sigma_1,\sigma_2,\sigma_3\ra$ where $\sigma_i=\up,\down$.}
\label{group.fig}
\end{figure}

To solve the Eq.~\eqref{Hcluster.eqn} first we choose a Y-shaped 4-site cluster $\Gamma$
shown in Fig.~\ref{group.fig}. The spin variables at every site have two possible states
giving a total of $2^4-16$ possible states for the cluster $\Gamma$.
Each state of this cluster is of the form $|\sigma_3,\sigma_2,\sigma_1,\sigma_0\ra$ where
$\sigma_a$ can take two possible values $\up,\down$ and the site indices $a=0,1,2,3$
are indicated in Fig.~\ref{group.fig}. The basis in this 16-dimensional
Hilbert space are as follows (for brevity we have dropped $|\ra$ from the representation
of basis states):
\begin{align}
&\left| 1 \right\rangle =\up \up \up \up~~  \left| 2 \right\rangle =\up \up \up \down~~ \left| 3 \right\rangle =\up \up \down \up~~ \left| 4 \right\rangle =\up \down \up \up \\
&\left| 5 \right\rangle =\down \up \up \up~~\left| 6 \right\rangle =\up \up \down \down ~~\left| 7 \right\rangle =\up \down \down \up~~\left| 8 \right\rangle =\down \down \up \up  \notag\\
&\left| 9 \right\rangle =\down \up \up \down~~\left| 10 \right\rangle =\up \down \up \down~~\left| 11 \right\rangle =\down \up \down \up~~\left| 12 \right\rangle =\up \down \down \down~~\notag\\
&\left| 13 \right\rangle =\down \up \down \down~~ \left| 14 \right\rangle =\down \down \up \down~~ \left| 15 \right\rangle =\down \down \down \up~~ \left| 16 \right\rangle =\down \down \down \down\nn
\end{align}
In the 4-site cluster of Fig.~\ref{group.fig} the positions $1,2,3$ are not nearest
neighbours of each other, while they are all neighbours of the site $0$. 
So the exchange interaction in the cluster takes place only between the site
$0$ and the above three sites. Hence the first term of Eq.~\eqref{Hcluster.eqn} 
for the 4-site cluster is,
\begin{eqnarray}
{{H}_{1}}=-\sum\limits_{\la a,b\ra\in \Gamma}{\tau _{a}^{x}\tau _{b}^{x}=-\left\{ \tau _{0}^{x}\tau _{1}^{x}+\tau _{0}^{x}\tau _{2}^{x}+\tau _{0}^{x}\tau _{3}^{x} \right\}}.
\end{eqnarray}
The effect of the above term on the bases is:
\begin{align}
&{{H}_{1}}\left| 1 \right\rangle =-\left\{ \left| 6 \right\rangle +\left| 9 \right\rangle +\left| 10 \right\rangle  \right\}\notag\\
&{{H}_{1}}\left| 2 \right\rangle =-\left\{ \left| 3 \right\rangle +\left| 4 \right\rangle +\left| 5 \right\rangle  \right\}\notag\\
&{{H}_{1}}\left| 3 \right\rangle =-\left\{ \left| 2 \right\rangle +\left| 12 \right\rangle +\left| 13 \right\rangle  \right\}\notag\\
&{{H}_{1}}\left| 4 \right\rangle =-\left\{ \left| 2 \right\rangle +\left| 12 \right\rangle +\left| 14 \right\rangle  \right\}\notag\\
&{{H}_{1}}\left| 5 \right\rangle =-\left\{ \left| 2 \right\rangle +\left| 13 \right\rangle +\left| 14 \right\rangle  \right\}\notag\\
&{{H}_{1}}\left| 6 \right\rangle =-\left\{ \left| 1 \right\rangle +\left| 7 \right\rangle +\left| 11 \right\rangle  \right\}\notag\\
&{{H}_{1}}\left| 7 \right\rangle =-\left\{ \left| 6 \right\rangle +\left| 10 \right\rangle +\left| 16 \right\rangle  \right\}\notag\\
&{{H}_{1}}\left| 8 \right\rangle =-\left\{ \left| 9 \right\rangle +\left| 10 \right\rangle +\left| 16 \right\rangle  \right\}\notag\\
&{{H}_{1}}\left| 9 \right\rangle =-\left\{ \left| 1 \right\rangle +\left| 8 \right\rangle +\left| 11 \right\rangle  \right\}\notag\\
&{{H}_{1}}\left| 10 \right\rangle =-\left\{ \left| 1 \right\rangle +\left| 7 \right\rangle +\left| 8 \right\rangle  \right\}\notag\\
&{{H}_{1}}\left| 11 \right\rangle =-\left\{ \left| 6 \right\rangle +\left| 9 \right\rangle +\left| 16 \right\rangle  \right\}\notag\\
&{{H}_{1}}\left| 12 \right\rangle =-\left\{ \left| 3 \right\rangle +\left| 4 \right\rangle +\left| 15 \right\rangle  \right\}\notag\\
&{{H}_{1}}\left| 13 \right\rangle =-\left\{ \left| 3 \right\rangle +\left| 5 \right\rangle +\left| 15 \right\rangle  \right\}\notag\\
&{{H}_{1}}\left| 14 \right\rangle =-\left\{ \left| 4 \right\rangle +\left| 5 \right\rangle +\left| 15 \right\rangle  \right\}\notag\\
&{{H}_{1}}\left| 15 \right\rangle =-\left\{ \left| 12 \right\rangle +\left| 13 \right\rangle +\left| 14 \right\rangle  \right\}\notag\\
&{{H}_{1}}\left| 16 \right\rangle =-\left\{ \left| 8 \right\rangle +\left| 11 \right\rangle +\left| 12 \right\rangle  \right\}.
\end{align}

The second term of the cluster Hamiltonian~\eqref{Hcluster.eqn} is
\begin{align}
{{H}_{2}}=-\frac{mz}{2}\sum_{a\in\Gamma}{\tau _{a}^{x}}=-\frac{mz}{2}\left\{ \tau _{0}^{x}+\tau _{1}^{x}+\tau _{2}^{x}+\tau _{3}^{x} \right\}
\end{align}
where $m$ is the Ising magnetization coupling the boundary sites of the cluster
to boundary sites of neighbouring clusters and
$z$ is the number of bonds connecting boundary sites $1,2,3$ to other clusters which for
this cluster is $z=2$. The effect of $H_2$ on the bases is given by,
\begin{align}
&{{H}_{2}}\left| 1 \right\rangle =-m\left\{ \left| 2 \right\rangle +\left| 3 \right\rangle +\left| 4 \right\rangle +\left| 5 \right\rangle  \right\}\nn\\
&{{H}_{2}}\left| 2 \right\rangle =-m\left\{ \left| 1 \right\rangle +\left| 6 \right\rangle +\left| 9 \right\rangle +\left| 10 \right\rangle  \right\}\nn\\
&{{H}_{2}}\left| 3 \right\rangle =-m\left\{ \left| 1 \right\rangle +\left| 6 \right\rangle +\left| 7 \right\rangle +\left| 11 \right\rangle  \right\}\nn\\
&{{H}_{2}}\left| 4 \right\rangle =-m\left\{ \left| 1 \right\rangle +\left| 7 \right\rangle +\left| 8 \right\rangle +\left| 10 \right\rangle  \right\}\nn\\
&{{H}_{2}}\left| 5 \right\rangle =-m\left\{ \left| 1 \right\rangle +\left| 8 \right\rangle +\left| 9 \right\rangle +\left| 11 \right\rangle  \right\}\nn\\
&{{H}_{2}}\left| 6 \right\rangle =-m\left\{ \left| 2 \right\rangle +\left| 3 \right\rangle +\left| 12 \right\rangle +\left| 13 \right\rangle  \right\}\nn\\
&{{H}_{2}}\left| 7 \right\rangle =-m\left\{ \left| 3 \right\rangle +\left| 4 \right\rangle +\left| 12 \right\rangle +\left| 15 \right\rangle  \right\}\nn\\
&{{H}_{2}}\left| 8 \right\rangle =-m\left\{ \left| 4 \right\rangle +\left| 5 \right\rangle +\left| 14 \right\rangle +\left| 15 \right\rangle  \right\}\nn\\
&{{H}_{2}}\left| 9 \right\rangle =-m\left\{ \left| 2 \right\rangle +\left| 5 \right\rangle +\left| 13 \right\rangle +\left| 14 \right\rangle  \right\}\nn\\
&{{H}_{2}}\left| 10 \right\rangle =-m\left\{ \left| 2 \right\rangle +\left| 4 \right\rangle +\left| 12 \right\rangle +\left| 14 \right\rangle  \right\}\nn\\
&{{H}_{2}}\left| 11 \right\rangle =-m\left\{ \left| 3 \right\rangle +\left| 5 \right\rangle +\left| 13 \right\rangle +\left| 15 \right\rangle  \right\}\nn\\
&{{H}_{2}}\left| 12 \right\rangle =-m\left\{ \left| 6 \right\rangle +\left| 7 \right\rangle +\left| 10 \right\rangle +\left| 16 \right\rangle  \right\}\nn\\
&{{H}_{2}}\left| 13 \right\rangle =-m\left\{ \left| 6 \right\rangle +\left| 9 \right\rangle +\left| 11 \right\rangle +\left| 16 \right\rangle  \right\}\nn\\
&{{H}_{2}}\left| 14 \right\rangle =-m\left\{ \left| 8 \right\rangle +\left| 9 \right\rangle +\left| 10 \right\rangle +\left| 16 \right\rangle  \right\}\nn\\
&{{H}_{2}}\left| 15 \right\rangle =-m\left\{ \left| 7 \right\rangle +\left| 8 \right\rangle +\left| 11 \right\rangle +\left| 16 \right\rangle  \right\}\nn\\
&{{H}_{2}}\left| 16 \right\rangle =-m\left\{ \left| 12 \right\rangle +\left| 13 \right\rangle +\left| 14 \right\rangle +\left| 15 \right\rangle  \right\}.
\end{align}
The last term of the cluster Hamiltonian~\eqref{Hcluster.eqn} is the transverse field term,
\begin{eqnarray}
{{H}_{3}}=h\sum\limits_{a\in\Gamma}{\tau _{a}^{z}}=h\left\{ \tau_0^z+\tau_1^z+\tau_2^z+\tau_3^z\right\}.
\end{eqnarray}
The above term acts on the $16$ bases as follows,
\begin{align}
&{{H}_{3}}\left| 1 \right\rangle =4h\left| 1 \right\rangle , {{H}_{3}}\left| 2 \right\rangle =2h\left| 2 \right\rangle \notag\\
&{{H}_{3}}\left| 3 \right\rangle =2h\left| 3 \right\rangle , {{H}_{3}}\left| 4 \right\rangle =2h\left| 4 \right\rangle \notag\\
&{{H}_{3}}\left| 5 \right\rangle =2h\left| 5 \right\rangle\notag\\
&{{H}_{3}}\left| 6 \right\rangle =0, {{H}_{3}}\left| 7 \right\rangle =0, {{H}_{3}}\left| 8 \right\rangle =0 \notag\\
&{{H}_{3}}\left| 9 \right\rangle =0, {{H}_{3}}\left| 10 \right\rangle =0, {{H}_{3}}\left| 11 \right\rangle =0 \notag\\
&{{H}_{3}}\left| 12 \right\rangle =-2h\left| 12 \right\rangle , {{H}_{3}}\left| 13 \right\rangle =-2h\left| 13 \right\rangle \notag\\
&{{H}_{3}}\left| 14 \right\rangle =-2h\left| 14 \right\rangle , {{H}_{3}}\left| 15 \right\rangle =-2h\left| 15 \right\rangle \notag\\
&{{H}_{3}}\left| 16 \right\rangle =-4h\left| 16 \right\rangle.
\end{align}

Let us proceed by employing symmetry considerations to reduce the
above 16-dimensional Hamiltonian to smaller blocks.
The Y-shaped cluster in Fig.~\ref{group.fig} is invariant under
rotations by $2\pi/3$ which is denoted by $C$ and the group of
rotation is formed by $\{C^0,C^1,C^2\}$. The effect of this 
operation on the site labels is 
\be
   C=\left\{ \begin{matrix}
   1\to 2  \\
   2\to 3  \\
   3\to 1  \\
\end{matrix} \right.\
\ee
Successive operations of $C$ on a prototypical state, e.g. $|12\ra$ gives 
the following pattern,
\begin{eqnarray}
|12\rangle\xrightarrow{C}|14\rangle\xrightarrow{C}|13\rangle\xrightarrow{C}|12\rangle
\end{eqnarray}
which is a concise representation of
\be
   C^0|12\ra=|12\ra,~~~C|12\ra = |14\ra,~~~C^2|12\ra=|13\ra
   \label{state12.eqn}
\ee
According to projection theorem of group theory a symmetry adopted
state in representation labeled by $n$ can be constructed from an arbitrary state
$|\phi\ra$ as,
\be
   |\psi^{(n)}\ra\sim \left(\sum_{g} g \Gamma_n[g]\right)|\phi\ra
\ee
where $g$ denotes member of the group, and $\Gamma_n(g)$ is the $n$'th irreducible
representation of element $g$ of the group. In the case of rotation group
the irreducible representations of the cyclic group are labeled by 
three integer (angular momenta) $n=0,\pm 1$ and are represented by
$\{\omega^0,\omega^n,\omega^{2n}\}$ where $\omega=\exp(2\pi i/3)$.
Compact way of expressing the above representations for the cyclic group is
$\Gamma_n(C^p)=\omega^{pn}$. 
This gives a symmetry adopted state build from e.g. basis state $|12\ra$ as
\be
   \left(C^0\omega^0+C^1\omega^n+C^2\omega^{2n}\right)|12\ra,
\ee
which after using~\eqref{state12.eqn} gives the following state
\be
   |12\ra+\omega^n|14\ra+\omega^{2n}|13\ra,
\ee
with definite discrete "angular momentum" $n$.

The same symmetry consideration could be applied to every other state
which is summarized as,
\begin{align}
&|6\rangle\xrightarrow{C}|10\rangle\xrightarrow{C}|9\rangle\xrightarrow{C}|6\rangle\notag\\
&|3\rangle\xrightarrow{C}|4\rangle\xrightarrow{C}|5\rangle\xrightarrow{C}|3\rangle\notag\\
&|7\rangle\xrightarrow{C}|8\rangle\xrightarrow{C}|11\rangle\xrightarrow{C}|7\rangle\notag\\
&|1\rangle\xrightarrow{C}|1\rangle,~~~~~|2\rangle\xrightarrow{C}|2\rangle\notag\\
&|15\rangle\xrightarrow{C}|15\rangle,~~~~~|16\rangle\xrightarrow{C}|16\rangle.
\end{align}
Let us now focus on the $n=+1$ sector. The $n=-1$ sector has identical spectrum by 
time-reversal symmetry. The $n=+1$ sector is spanned by the following normalized states,
\begin{align}
&\left| {{\alpha }_{1}} \right\rangle =\frac{1}{\sqrt{3}}\left( \left| 12 \right\rangle +\omega \left| 14 \right\rangle +{{\omega }^{2}}\left| 13 \right\rangle  \right) \nn\\
&\left| {{\alpha }_{2}} \right\rangle =\frac{1}{\sqrt{3}}\left( \left| 3 \right\rangle +\omega \left| 4 \right\rangle +{{\omega }^{2}}\left| 5 \right\rangle  \right) \nn\\
&\left| {{\alpha }_{3}} \right\rangle =\frac{1}{\sqrt{3}}\left( \left| 6 \right\rangle +\omega \left| 10 \right\rangle +{{\omega }^{2}}\left| 9 \right\rangle  \right) \nn\\
&\left| {{\alpha }_{4}} \right\rangle =\frac{1}{\sqrt{3}}\left( \left| 7 \right\rangle +\omega \left| 8 \right\rangle +{{\omega }^{2}}\left| 11 \right\rangle  \right)
\end{align}
where $\omega=e^{2i\pi/3}$. The first term of the cluster Hamiltonian on the
above states has the following effect:
\begin{eqnarray}
{{H}_{1}}\left| {{\alpha }_{1}} \right\rangle =-\Omega_-\left| {{\alpha }_{2}} \right\rangle\notag\\
{{H}_{1}}\left| {{\alpha }_{2}} \right\rangle =-\Omega_+\left| {{\alpha }_{1}} \right\rangle \notag\\
{{H}_{1}}\left| {{\alpha }_{3}} \right\rangle =-\Omega_+\left| {{\alpha }_{4}} \right\rangle \notag\\
{{H}_{1}}\left| {{\alpha }_{4}} \right\rangle =-\Omega_-\left| {{\alpha }_{3}} \right\rangle
\end{eqnarray}
where $\Omega_\pm=1+\omega^{\pm 1}=\exp(\pm i\pi/3)$. The matrix elements
$\la \alpha_i|H_1\alpha_j\ra$ are represented in matrix form as,
\begin{equation}
{{\tilde{H}}_{1}}=-\left( \begin{matrix}
   0 & \Omega_- & 0 & 0  \\
   \Omega_+ & 0 & 0 & 0  \\
   0 & 0 & 0 & \Omega_+   \\
   0 & 0 & \Omega_- & 0  
\end{matrix} \right)
\end{equation}
Similarly the $H_2$ term in this basis becomes,
\begin{eqnarray}
\tilde{H}_2=-m\left( \begin{matrix}
   0 & 0 & \Omega_- & 1   \\
   0 & 0 & 1 & \Omega_+   \\
   \Omega_+ & 1 & 0 & 0  \\
   1 & \Omega_- & 0 & 0 
\end{matrix} \right)
\end{eqnarray}
The transverse field term was already diagonal in the original basis,
and remains so in the symmetry adopted basis for the $n=+1$ sector,
\begin{eqnarray}
{{\tilde{H}}_{3}}=-\left( \begin{matrix}
   2h & {} & {} & {}  \\
   {} & -2h & {} & {}  \\
   {} & {} & 0 & {}  \\
   {} & {} & {} & 0  
\end{matrix} \right)
\end{eqnarray}
Adding up the above matrices we obtain the matrix representation 
of the ITF Hamiltonian for Y-shaped cluster in the $n=+1$ sector as,
\begin{eqnarray}
&&{{\tilde{H}}}={{\tilde{H}}_{1}}+{{\tilde{H}}_{2}}+
{{\tilde{H}}_{3}}=\notag\\
&& -\left( \begin{matrix}
   2h & {{e}^{{i\pi }/{3}}} & m{{e}^{{i\pi }/{3}}} & m  \\
   {{e}^{{-i\pi }/{3}}} & -2h & m & m{{e}^{{-i\pi }/{3}}}  \\
   m{{e}^{{-i\pi }/{3}}} & m & 0 & {{e}^{{-i\pi }/{3}}}  \\
   m & m{{e}^{{i\pi }/{3}}} & {{e}^{{i\pi }/{3}}} & 0  \\
\end{matrix} \right)
\end{eqnarray}
The matrix representation for the $n=-1$ sector is simply obtained from 
the above equation by complex conjugation $i\to -i$ corresponding to 
time reversal operation. The matrix representation in the $n=0$ sector 
can be constructed in similar way. The ground state is the least eigen-value
among all sectors with various $n$ values. Here it turns out that the ground
state belongs to $n=0$ sector.

\subsection{Hexagonal cluster}
The details of the group theory consideration for larger clusters is
similar to Y-shaped cluster. In this section for reference we only
provide explicit representation of all $2^6=64$ basis states and
the effects of cluster ITF Hamiltonian on it. As can be seen in Fig.~\ref{boththem.fig}
each hexagonal cluster $\Gamma$ is connected to the rest of the lattice
by $z=1$ neighbour. The basis is labeled as before by 
$|\sigma_5,\sigma_4,\sigma_3,\sigma_2,\sigma_1,\sigma_0\ra$ where the site index
$a$ in $\sigma_a$ varies from $0$ to $5$ as depicted in right panel of Fig.~\ref{boththem.fig}.

\begin{align}
& \left| 1 \right\rangle =\down \down \down \down \down \down~~~~\left| 2 \right\rangle =\down \down \down \down \down \up~~~~\left| 3 \right\rangle =\down \down \down \down \up \down~~~~\left| 4 \right\rangle =\down \down \down \down \up \up  \nn\\
& \left| 5 \right\rangle =\down \down \down \up \down \down~~~~\left| 6 \right\rangle =\down \down \down \up \down \up~~~~\left| 7 \right\rangle =\down \down \down \up \up \down~~~~\left| 8 \right\rangle =\down \down \down \up \up \up  \nn\\
& \left| 9 \right\rangle =\down \down \up \down \down \down~~~~\left| 10 \right\rangle =\down \down \up \down \down \up~~\left| 11 \right\rangle =\down \down \up \down \up \down~~\left| 12 \right\rangle =\down \down \up \down \up \up  \nn\\
& \left| 13 \right\rangle =\down \down \up \up \down \down~~\left| 14 \right\rangle =\down \down \up \up \down \up ~~\left| 15 \right\rangle =\down \down \up \up \up \down ~~\left| 16 \right\rangle =\down \down \up \up \up \up  \nn\\
& \left| 17 \right\rangle =\down \up \down \down \down \down ~~\left| 18 \right\rangle =\down \up \down \down \down \up ~~\left| 19 \right\rangle =\down \up \down \down \up \down ~~\left| 20 \right\rangle =\down \up \down \down \up \up  \nn\\
& \left| 21 \right\rangle =\down \up \down \up \down \down ~~\left| 22 \right\rangle =\down \up \down \up \down \up ~~\left| 23 \right\rangle =\down \up \down \up \up \down ~~\left| 24 \right\rangle =\down \up \down \up \up \up  \nn\\
& \left| 25 \right\rangle =\down \up \up \down \down \down ~~\left| 26 \right\rangle =\down \up \up \down \down \up ~~\left| 27 \right\rangle =\down \up \up \down \up \down ~~\left| 28 \right\rangle =\down \up \up \down \up \up  \nn\\
& \left| 29 \right\rangle =\down \up \up \up \down \down ~~\left| 30 \right\rangle =\down \up \up \up \down \up ~~\left| 31 \right\rangle =\down \up \up \up \up \down ~~\left| 32 \right\rangle =\down \up \up \up \up \up  \nn\\
& \left| 33 \right\rangle =\up \down \down \down \down \down ~~\left| 34 \right\rangle =\up \down \down \down \down \up ~~\left| 35 \right\rangle =\up \down \down \down \up \down ~~\left| 36 \right\rangle =\up \down \down \down \up \up  \nn\\
& \left| 37 \right\rangle =\up \down \down \up \down \down ~~\left| 38 \right\rangle =\up \down \down \up \down \up~~\left| 39 \right\rangle =\up \down \down \up \up \down~~\left| 40 \right\rangle =\up \down \down \up \up \up  \nn\\
& \left| 41 \right\rangle =\up \down \up \down \down \down~~\left| 42 \right\rangle =\up \down \up \down \down \up~~\left| 43 \right\rangle =\up \down \up \down \up \down~~\left| 44 \right\rangle =\up \down \up \down \up \up  \nn\\
& \left| 45 \right\rangle =\up \down \up \up \down \down~~\left| 46 \right\rangle =\up \down \up \up \down \up~~\left| 47 \right\rangle =\up \down \up \up \up \down~~\left| 48 \right\rangle =\up \down \up \up \up \up  \nn\\
& \left| 49 \right\rangle =\up \up \down \down \down \down~~\left| 50 \right\rangle =\up \up \down \down \down \up~~\left| 51 \right\rangle =\up \up \down \down \up \down~~\left| 52 \right\rangle =\up \up \down \down \up \up  \nn\\
& \left| 53 \right\rangle =\up \up \down \up \down \down~~\left| 54 \right\rangle =\up \up \down \up \down \up~~\left| 55 \right\rangle =\up \up \down \up \up \down~~\left| 56 \right\rangle =\up \up \down \up \up \up  \nn\\
& \left| 57 \right\rangle =\up \up \up \down \down \down~~\left| 58 \right\rangle =\up \up \up \down \down \up~~\left| 59 \right\rangle =\up \up \up \down \up \down~~\left| 60 \right\rangle =\up \up \up \down \up \up  \nn\\
& \left| 61 \right\rangle =\up \up \up \up \down \down~~\left| 62 \right\rangle =\up \up \up \up \down \up~~\left| 63 \right\rangle =\up \up \up \up \up \down~~\left| 64 \right\rangle =\up \up \up \up \up \up
\end{align}
the first term of Eq.~\eqref{Hcluster.eqn} for $6$-site cluster is,
\begin{align}
   {{H}_{1}}&=-\sum_{\la a,b\ra\in \Gamma} \tau^x_a \tau^x_b \\
	    &-\left( \tau _{0}^{x}\tau _{1}^{x}+ \tau _{1}^{x}\tau _{2}^{x}+\tau _{2}^{x}\tau _{3}^{x}+\tau _{3}^{x}\tau _{4}^{x}+\tau _{4}^{x}\tau _{5}^{x}+\tau _{5}^{x}\tau _{0}^{x}\right).\nn
\end{align}
The effect of the above term on the $64$-basis states is,

\begin{align}
&{{H}_{1}}\left| 1 \right\rangle =-\left\{ \left| 4 \right\rangle +\left| 7 \right\rangle +\left| 13 \right\rangle +\left| 25 \right\rangle +\left| 34 \right\rangle +\left| 49 \right\rangle  \right\}\notag\\
&{{H}_{1}}\left| 2 \right\rangle =-\left\{ \left| 3 \right\rangle +\left| 8 \right\rangle +\left| 14 \right\rangle +\left| 26 \right\rangle +\left| 33 \right\rangle +\left| 50 \right\rangle  \right\}\notag\\
&{{H}_{1}}\left| 3 \right\rangle =-\left\{ \left| 2 \right\rangle +\left| 5 \right\rangle +\left| 15 \right\rangle +\left| 27 \right\rangle +\left| 36 \right\rangle +\left| 51 \right\rangle  \right\}\notag\\
&{{H}_{1}}\left| 4 \right\rangle =-\left\{ \left| 1 \right\rangle +\left| 6 \right\rangle +\left| 16 \right\rangle +\left| 28 \right\rangle +\left| 35 \right\rangle +\left| 52 \right\rangle  \right\}\notag\\
&{{H}_{1}}\left| 5 \right\rangle =-\left\{ \left| 8 \right\rangle +\left| 3 \right\rangle +\left| 9 \right\rangle +\left| 29 \right\rangle +\left| 38 \right\rangle +\left| 53 \right\rangle  \right\}\notag\\
&{{H}_{1}}\left| 6 \right\rangle =-\left\{ \left| 7 \right\rangle +\left| 4 \right\rangle +\left| 10 \right\rangle +\left| 30 \right\rangle +\left| 37 \right\rangle +\left| 54 \right\rangle  \right\}\notag\\
&{{H}_{1}}\left| 7 \right\rangle =-\left\{ \left| 6 \right\rangle +\left| 1 \right\rangle +\left| 11 \right\rangle +\left| 31 \right\rangle +\left| 40 \right\rangle +\left| 55 \right\rangle  \right\}\notag\\
&{{H}_{1}}\left| 8 \right\rangle =-\left\{ \left| 5 \right\rangle +\left| 2 \right\rangle +\left| 12 \right\rangle +\left| 32 \right\rangle +\left| 39 \right\rangle +\left| 56 \right\rangle  \right\}\notag\\
&{{H}_{1}}\left| 9 \right\rangle =-\left\{ \left| 12 \right\rangle +\left| 15 \right\rangle +\left| 5 \right\rangle +\left| 17 \right\rangle +\left| 42 \right\rangle +\left| 57 \right\rangle  \right\}\notag\\
&{{H}_{1}}\left| 10 \right\rangle =-\left\{ \left| 11 \right\rangle +\left| 16 \right\rangle +\left| 6 \right\rangle +\left| 18 \right\rangle +\left| 41 \right\rangle +\left| 58 \right\rangle  \right\}\notag
\end{align}
\begin{align}
&{{H}_{1}}\left| 11 \right\rangle =-\left\{ \left| 10 \right\rangle +\left| 13 \right\rangle +\left| 7 \right\rangle +\left| 19 \right\rangle +\left| 44 \right\rangle +\left| 59 \right\rangle  \right\}\notag\\
&{{H}_{1}}\left| 12 \right\rangle =-\left\{ \left| 9 \right\rangle +\left| 14 \right\rangle +\left| 8 \right\rangle +\left| 20 \right\rangle +\left| 43 \right\rangle +\left| 60 \right\rangle  \right\}\notag\\
&{{H}_{1}}\left| 13 \right\rangle =-\left\{ \left| 16 \right\rangle +\left| 11 \right\rangle +\left| 1 \right\rangle +\left| 21 \right\rangle +\left| 46 \right\rangle +\left| 61 \right\rangle  \right\}\notag\\
&{{H}_{1}}\left| 14 \right\rangle =-\left\{ \left| 15 \right\rangle +\left| 12 \right\rangle +\left| 2 \right\rangle +\left| 22 \right\rangle +\left| 45 \right\rangle +\left| 62 \right\rangle  \right\}\notag\\
&{{H}_{1}}\left| 15 \right\rangle =-\left\{ \left| 14 \right\rangle +\left| 9 \right\rangle +\left| 3 \right\rangle +\left| 23 \right\rangle +\left| 48 \right\rangle +\left| 63 \right\rangle  \right\}\notag\\
&{{H}_{1}}\left| 16 \right\rangle =-\left\{ \left| 13 \right\rangle +\left| 10 \right\rangle +\left| 4 \right\rangle +\left| 24 \right\rangle +\left| 47 \right\rangle +\left| 64 \right\rangle  \right\}\notag\\
&{{H}_{1}}\left| 17 \right\rangle =-\left\{ \left| 20 \right\rangle +\left| 23 \right\rangle +\left| 29 \right\rangle +\left| 9 \right\rangle +\left| 50 \right\rangle +\left| 33 \right\rangle  \right\}\notag\\
&{{H}_{1}}\left| 18 \right\rangle =-\left\{ \left| 19 \right\rangle +\left| 24 \right\rangle +\left| 30 \right\rangle +\left| 10 \right\rangle +\left| 49 \right\rangle +\left| 34 \right\rangle  \right\}\notag\\
&{{H}_{1}}\left| 19 \right\rangle =-\left\{ \left| 18 \right\rangle +\left| 21 \right\rangle +\left| 31 \right\rangle +\left| 11 \right\rangle +\left| 52 \right\rangle +\left| 35 \right\rangle  \right\}\notag\\
&{{H}_{1}}\left| 20 \right\rangle =-\left\{ \left| 17 \right\rangle +\left| 22 \right\rangle +\left| 32 \right\rangle +\left| 12 \right\rangle +\left| 51 \right\rangle +\left| 36 \right\rangle  \right\}\notag
\end{align}
\begin{align}
&{{H}_{1}}\left| 21 \right\rangle =-\left\{ \left| 24 \right\rangle +\left| 19 \right\rangle +\left| 25 \right\rangle +\left| 13 \right\rangle +\left| 54 \right\rangle +\left| 37 \right\rangle  \right\}\notag\\
&{{H}_{1}}\left| 22 \right\rangle =-\left\{ \left| 23 \right\rangle +\left| 20 \right\rangle +\left| 26 \right\rangle +\left| 14 \right\rangle +\left| 53 \right\rangle +\left| 38 \right\rangle  \right\}\notag\\
&{{H}_{1}}\left| 23 \right\rangle =-\left\{ \left| 22 \right\rangle +\left| 17 \right\rangle +\left| 27 \right\rangle +\left| 15 \right\rangle +\left| 56 \right\rangle +\left| 39 \right\rangle  \right\}\notag\\
&{{H}_{1}}\left| 24 \right\rangle =-\left\{ \left| 21 \right\rangle +\left| 18 \right\rangle +\left| 28 \right\rangle +\left| 16 \right\rangle +\left| 55 \right\rangle +\left| 40 \right\rangle  \right\}\notag\\
&{{H}_{1}}\left| 25 \right\rangle =-\left\{ \left| 28 \right\rangle +\left| 15 \right\rangle +\left| 21 \right\rangle +\left| 1 \right\rangle +\left| 58 \right\rangle +\left| 41 \right\rangle  \right\}\notag\\
&{{H}_{1}}\left| 26 \right\rangle =-\left\{ \left| 27 \right\rangle +\left| 16 \right\rangle +\left| 22 \right\rangle +\left| 2 \right\rangle +\left| 57 \right\rangle +\left| 42 \right\rangle  \right\}\notag\\
&{{H}_{1}}\left| 27 \right\rangle =-\left\{ \left| 26 \right\rangle +\left| 29 \right\rangle +\left| 23 \right\rangle +\left| 3 \right\rangle +\left| 60 \right\rangle +\left| 43 \right\rangle  \right\}\notag\\
&{{H}_{1}}\left| 28 \right\rangle =-\left\{ \left| 25 \right\rangle +\left| 30 \right\rangle +\left| 24 \right\rangle +\left| 4 \right\rangle +\left| 59 \right\rangle +\left| 44 \right\rangle  \right\}\notag\\
&{{H}_{1}}\left| 29 \right\rangle =-\left\{ \left| 32 \right\rangle +\left| 27 \right\rangle +\left| 17 \right\rangle +\left| 5 \right\rangle +\left| 62 \right\rangle +\left| 45 \right\rangle  \right\}\notag\\
&{{H}_{1}}\left| 30 \right\rangle =-\left\{ \left| 31 \right\rangle +\left| 28 \right\rangle +\left| 18 \right\rangle +\left| 6 \right\rangle +\left| 61 \right\rangle +\left| 46 \right\rangle  \right\}\notag
\end{align}
\begin{align}
&{{H}_{1}}\left| 31 \right\rangle =-\left\{ \left| 30 \right\rangle +\left| 25 \right\rangle +\left| 19 \right\rangle +\left| 7 \right\rangle +\left| 64 \right\rangle +\left| 47 \right\rangle  \right\}\notag\\
&{{H}_{1}}\left| 32 \right\rangle =-\left\{ \left| 29 \right\rangle +\left| 26 \right\rangle +\left| 20 \right\rangle +\left| 8 \right\rangle +\left| 63 \right\rangle +\left| 48 \right\rangle  \right\}\notag\\
&{{H}_{1}}\left| 33 \right\rangle =-\left\{ \left| 36 \right\rangle +\left| 39 \right\rangle +\left| 45 \right\rangle +\left| 57 \right\rangle +\left| 2 \right\rangle +\left| 17 \right\rangle  \right\}\notag\\
&{{H}_{1}}\left| 34 \right\rangle =-\left\{ \left| 35 \right\rangle +\left| 40 \right\rangle +\left| 46 \right\rangle +\left| 58 \right\rangle +\left| 1 \right\rangle +\left| 18 \right\rangle  \right\}\notag\\
&{{H}_{1}}\left| 35 \right\rangle =-\left\{ \left| 34 \right\rangle +\left| 37 \right\rangle +\left| 47 \right\rangle +\left| 59 \right\rangle +\left| 4 \right\rangle +\left| 19 \right\rangle  \right\}\notag\\
&{{H}_{1}}\left| 36 \right\rangle =-\left\{ \left| 33 \right\rangle +\left| 38 \right\rangle +\left| 48 \right\rangle +\left| 60 \right\rangle +\left| 3 \right\rangle +\left| 20 \right\rangle  \right\}\notag\\
&{{H}_{1}}\left| 37 \right\rangle =-\left\{ \left| 40 \right\rangle +\left| 35 \right\rangle +\left| 41 \right\rangle +\left| 61 \right\rangle +\left| 6 \right\rangle +\left| 21 \right\rangle  \right\}\notag\\
&{{H}_{1}}\left| 38 \right\rangle =-\left\{ \left| 39 \right\rangle +\left| 36 \right\rangle +\left| 42 \right\rangle +\left| 62 \right\rangle +\left| 5 \right\rangle +\left| 22 \right\rangle  \right\}\notag\\
&{{H}_{1}}\left| 39 \right\rangle =-\left\{ \left| 38 \right\rangle +\left| 33 \right\rangle +\left| 43 \right\rangle +\left| 63 \right\rangle +\left| 8 \right\rangle +\left| 23 \right\rangle  \right\}\notag\\
&{{H}_{1}}\left| 40 \right\rangle =-\left\{ \left| 37 \right\rangle +\left| 34 \right\rangle +\left| 44 \right\rangle +\left| 64 \right\rangle +\left| 7 \right\rangle +\left| 24 \right\rangle  \right\}\notag
\end{align}
\begin{align}
&{{H}_{1}}\left| 41 \right\rangle =-\left\{ \left| 44 \right\rangle +\left| 47 \right\rangle +\left| 37 \right\rangle +\left| 49 \right\rangle +\left| 10 \right\rangle +\left| 25 \right\rangle  \right\}\notag\\
&{{H}_{1}}\left| 42 \right\rangle =-\left\{ \left| 43 \right\rangle +\left| 48 \right\rangle +\left| 38 \right\rangle +\left| 50 \right\rangle +\left| 9 \right\rangle +\left| 26 \right\rangle  \right\}\notag\\
&{{H}_{1}}\left| 43 \right\rangle =-\left\{ \left| 42 \right\rangle +\left| 45 \right\rangle +\left| 39 \right\rangle +\left| 51 \right\rangle +\left| 12 \right\rangle +\left| 27 \right\rangle  \right\}\notag\\
&{{H}_{1}}\left| 44 \right\rangle =-\left\{ \left| 41 \right\rangle +\left| 46 \right\rangle +\left| 40 \right\rangle +\left| 52 \right\rangle +\left| 11 \right\rangle +\left| 28 \right\rangle  \right\}\notag\\
&{{H}_{1}}\left| 45 \right\rangle =-\left\{ \left| 48 \right\rangle +\left| 43 \right\rangle +\left| 33 \right\rangle +\left| 53 \right\rangle +\left| 14 \right\rangle +\left| 29 \right\rangle  \right\}\notag\\
&{{H}_{1}}\left| 46 \right\rangle =-\left\{ \left| 47 \right\rangle +\left| 44 \right\rangle +\left| 34 \right\rangle +\left| 54 \right\rangle +\left| 13 \right\rangle +\left| 30 \right\rangle  \right\}\notag\\
&{{H}_{1}}\left| 47 \right\rangle =-\left\{ \left| 46 \right\rangle +\left| 41 \right\rangle +\left| 35 \right\rangle +\left| 55 \right\rangle +\left| 16 \right\rangle +\left| 31 \right\rangle  \right\}\notag\\
&{{H}_{1}}\left| 48 \right\rangle =-\left\{ \left| 45 \right\rangle +\left| 42 \right\rangle +\left| 36 \right\rangle +\left| 56 \right\rangle +\left| 15 \right\rangle +\left| 32 \right\rangle  \right\}\notag\\
&{{H}_{1}}\left| 49 \right\rangle =-\left\{ \left| 52 \right\rangle +\left| 55 \right\rangle +\left| 61 \right\rangle +\left| 41 \right\rangle +\left| 18 \right\rangle +\left| 1 \right\rangle  \right\}\notag\\
&{{H}_{1}}\left| 50 \right\rangle =-\left\{ \left| 51 \right\rangle +\left| 56 \right\rangle +\left| 62 \right\rangle +\left| 42 \right\rangle +\left| 17 \right\rangle +\left| 2 \right\rangle  \right\}\notag
\end{align}
\begin{align}
&{{H}_{1}}\left| 51 \right\rangle =-\left\{ \left| 50 \right\rangle +\left| 53 \right\rangle +\left| 63 \right\rangle +\left| 43 \right\rangle +\left| 20 \right\rangle +\left| 3 \right\rangle  \right\}\notag\\
&{{H}_{1}}\left| 52 \right\rangle =-\left\{ \left| 49 \right\rangle +\left| 54 \right\rangle +\left| 64 \right\rangle +\left| 44 \right\rangle +\left| 19 \right\rangle +\left| 4 \right\rangle  \right\}\notag\\
&{{H}_{1}}\left| 53 \right\rangle =-\left\{ \left| 56 \right\rangle +\left| 51 \right\rangle +\left| 57 \right\rangle +\left| 45 \right\rangle +\left| 22 \right\rangle +\left| 5 \right\rangle  \right\}\notag\\
&{{H}_{1}}\left| 54 \right\rangle =-\left\{ \left| 55 \right\rangle +\left| 52 \right\rangle +\left| 58 \right\rangle +\left| 46 \right\rangle +\left| 21 \right\rangle +\left| 6 \right\rangle  \right\}\notag\\
&{{H}_{1}}\left| 55 \right\rangle =-\left\{ \left| 54 \right\rangle +\left| 49 \right\rangle +\left| 59 \right\rangle +\left| 47 \right\rangle +\left| 24 \right\rangle +\left| 7 \right\rangle  \right\}\notag\\
&{{H}_{1}}\left| 56 \right\rangle =-\left\{ \left| 53 \right\rangle +\left| 50 \right\rangle +\left| 60 \right\rangle +\left| 48 \right\rangle +\left| 23 \right\rangle +\left| 8 \right\rangle  \right\}\notag\\
&{{H}_{1}}\left| 57 \right\rangle =-\left\{ \left| 60 \right\rangle +\left| 63 \right\rangle +\left| 53 \right\rangle +\left| 33 \right\rangle +\left| 26 \right\rangle +\left| 9 \right\rangle  \right\}\notag\\
&{{H}_{1}}\left| 58 \right\rangle =-\left\{ \left| 59 \right\rangle +\left| 64 \right\rangle +\left| 54 \right\rangle +\left| 34 \right\rangle +\left| 25 \right\rangle +\left| 10 \right\rangle  \right\}\notag\\
&{{H}_{1}}\left| 59 \right\rangle =-\left\{ \left| 58 \right\rangle +\left| 61 \right\rangle +\left| 55 \right\rangle +\left| 35 \right\rangle +\left| 28 \right\rangle +\left| 11 \right\rangle  \right\}\notag\\
&{{H}_{1}}\left| 60 \right\rangle =-\left\{ \left| 57 \right\rangle +\left| 62 \right\rangle +\left| 56 \right\rangle +\left| 36 \right\rangle +\left| 27 \right\rangle +\left| 12 \right\rangle  \right\}\notag
\end{align}
\begin{align}
&{{H}_{1}}\left| 61 \right\rangle =-\left\{ \left| 64 \right\rangle +\left| 59 \right\rangle +\left| 49 \right\rangle +\left| 37 \right\rangle +\left| 30 \right\rangle +\left| 13 \right\rangle  \right\}\notag\\
&{{H}_{1}}\left| 62 \right\rangle =-\left\{ \left| 63 \right\rangle +\left| 60 \right\rangle +\left| 50 \right\rangle +\left| 38 \right\rangle +\left| 29 \right\rangle +\left| 14 \right\rangle  \right\}\notag\\
&{{H}_{1}}\left| 63 \right\rangle =-\left\{ \left| 62 \right\rangle +\left| 57 \right\rangle +\left| 51 \right\rangle +\left| 39 \right\rangle +\left| 32 \right\rangle +\left| 15 \right\rangle  \right\}\notag\\
&{{H}_{1}}\left| 64 \right\rangle =-\left\{ \left| 61 \right\rangle +\left| 58 \right\rangle +\left| 52 \right\rangle +\left| 40 \right\rangle +\left| 31 \right\rangle +\left| 16 \right\rangle  \right\}\notag\\
\end{align}
The second term of the cluster Hamiltonian~\eqref{Hcluster.eqn} is
\begin{align}
{{H}_{2}}=-\frac{m}{2}\left\{\tau_0^x+\tau_1^x+\tau_2^x+\tau_3^x+\tau_4^x+\tau_5^x \right\}
\end{align}
for the above equation the number of bonds connecting to other clusters is $z=1$. The effect of $H_2$ on the basis is given by:
\begin{align}
&{{H}_{2}}\left| 1 \right\rangle =\frac{-m}{2}\left\{ \left| 2 \right\rangle +\left| 3 \right\rangle +\left| 5 \right\rangle +\left| 9 \right\rangle +\left| 17 \right\rangle +\left| 33 \right\rangle  \right\}\notag\\
&{{H}_{2}}\left| 2 \right\rangle =\frac{-m}{2}\left\{ \left| 1 \right\rangle +\left| 4 \right\rangle +\left| 6 \right\rangle +\left| 10 \right\rangle +\left| 18 \right\rangle +\left| 34 \right\rangle  \right\}\notag\\
&{{H}_{2}}\left| 3 \right\rangle =\frac{-m}{2}\left\{ \left| 4 \right\rangle +\left| 1 \right\rangle +\left| 7 \right\rangle +\left| 11 \right\rangle +\left| 19 \right\rangle +\left| 35 \right\rangle  \right\}\notag\\
&{{H}_{2}}\left| 4 \right\rangle =\frac{-m}{2}\left\{ \left| 3 \right\rangle +\left| 2 \right\rangle +\left| 8 \right\rangle +\left| 12 \right\rangle +\left| 20 \right\rangle +\left| 36 \right\rangle  \right\}\notag\\
&{{H}_{2}}\left| 5 \right\rangle =\frac{-m}{2}\left\{ \left| 6 \right\rangle +\left| 7 \right\rangle +\left| 1 \right\rangle +\left| 13 \right\rangle +\left| 21 \right\rangle +\left| 37 \right\rangle  \right\}\notag\\
&{{H}_{2}}\left| 6 \right\rangle =\frac{-m}{2}\left\{ \left| 5 \right\rangle +\left| 8 \right\rangle +\left| 2 \right\rangle +\left| 14 \right\rangle +\left| 22 \right\rangle +\left| 38 \right\rangle  \right\}\notag\\
&{{H}_{2}}\left| 7 \right\rangle =\frac{-m}{2}\left\{ \left| 8 \right\rangle +\left| 5 \right\rangle +\left| 3 \right\rangle +\left| 15 \right\rangle +\left| 23 \right\rangle +\left| 39 \right\rangle  \right\}\notag\\
&{{H}_{2}}\left| 8 \right\rangle =\frac{-m}{2}\left\{ \left| 7 \right\rangle +\left| 6 \right\rangle +\left| 4 \right\rangle +\left| 16 \right\rangle +\left| 24 \right\rangle +\left| 40 \right\rangle  \right\}\notag\\
&{{H}_{2}}\left| 9 \right\rangle =\frac{-m}{2}\left\{ \left| 10 \right\rangle +\left| 11 \right\rangle +\left| 13 \right\rangle +\left| 1 \right\rangle +\left| 25 \right\rangle +\left| 41 \right\rangle  \right\}\notag\\
&{{H}_{2}}\left| 10 \right\rangle =\frac{-m}{2}\left\{ \left| 9 \right\rangle +\left| 12 \right\rangle +\left| 14 \right\rangle +\left| 2 \right\rangle +\left| 26 \right\rangle +\left| 42 \right\rangle  \right\}\notag
\end{align}
\begin{align}
&{{H}_{2}}\left| 11 \right\rangle =\frac{-m}{2}\left\{ \left| 12 \right\rangle +\left| 9 \right\rangle +\left| 15 \right\rangle +\left| 3 \right\rangle +\left| 27 \right\rangle +\left| 43 \right\rangle  \right\}\notag\\
&{{H}_{2}}\left| 12 \right\rangle =\frac{-m}{2}\left\{ \left| 11 \right\rangle +\left| 10 \right\rangle +\left| 16 \right\rangle +\left| 4 \right\rangle +\left| 28 \right\rangle +\left| 44 \right\rangle  \right\}\notag\\
&{{H}_{2}}\left| 13 \right\rangle =\frac{-m}{2}\left\{ \left| 14 \right\rangle +\left| 15 \right\rangle +\left| 9 \right\rangle +\left| 5 \right\rangle +\left| 29 \right\rangle +\left| 45 \right\rangle  \right\}\notag\\
&{{H}_{2}}\left| 14 \right\rangle =\frac{-m}{2}\left\{ \left| 13 \right\rangle +\left| 16 \right\rangle +\left| 10 \right\rangle +\left| 6 \right\rangle +\left| 30 \right\rangle +\left| 46 \right\rangle  \right\}\notag\\
&{{H}_{2}}\left| 15 \right\rangle =\frac{-m}{2}\left\{ \left| 16 \right\rangle +\left| 13 \right\rangle +\left| 11 \right\rangle +\left| 7 \right\rangle +\left| 31 \right\rangle +\left| 47 \right\rangle  \right\}\notag\\
&{{H}_{2}}\left| 16 \right\rangle =\frac{-m}{2}\left\{ \left| 15 \right\rangle +\left| 14 \right\rangle +\left| 12 \right\rangle +\left| 8 \right\rangle +\left| 32 \right\rangle +\left| 48 \right\rangle  \right\}\notag\\
&{{H}_{2}}\left| 17 \right\rangle =\frac{-m}{2}\left\{ \left| 18 \right\rangle +\left| 19 \right\rangle +\left| 21 \right\rangle +\left| 25 \right\rangle +\left| 1 \right\rangle +\left| 49 \right\rangle  \right\}\notag\\
&{{H}_{2}}\left| 18 \right\rangle =\frac{-m}{2}\left\{ \left| 17 \right\rangle +\left| 20 \right\rangle +\left| 22 \right\rangle +\left| 26 \right\rangle +\left| 2 \right\rangle +\left| 50 \right\rangle  \right\}\notag\\
&{{H}_{2}}\left| 19 \right\rangle =\frac{-m}{2}\left\{ \left| 20 \right\rangle +\left| 17 \right\rangle +\left| 23 \right\rangle +\left| 27 \right\rangle +\left| 3 \right\rangle +\left| 51 \right\rangle  \right\}\notag\\
&{{H}_{2}}\left| 20 \right\rangle =\frac{-m}{2}\left\{ \left| 19 \right\rangle +\left| 18 \right\rangle +\left| 24 \right\rangle +\left| 28 \right\rangle +\left| 4 \right\rangle +\left| 52 \right\rangle  \right\}\notag
\end{align}
\begin{align}
&{{H}_{2}}\left| 21 \right\rangle =\frac{-m}{2}\left\{ \left| 22 \right\rangle +\left| 23 \right\rangle +\left| 17 \right\rangle +\left| 29 \right\rangle +\left| 5 \right\rangle +\left| 53 \right\rangle  \right\}\notag\\
&{{H}_{2}}\left| 22 \right\rangle =\frac{-m}{2}\left\{ \left| 21 \right\rangle +\left| 24 \right\rangle +\left| 18 \right\rangle +\left| 30 \right\rangle +\left| 6 \right\rangle +\left| 54 \right\rangle  \right\}\notag\\
&{{H}_{2}}\left| 23 \right\rangle =\frac{-m}{2}\left\{ \left| 24 \right\rangle +\left| 21 \right\rangle +\left| 19 \right\rangle +\left| 31 \right\rangle +\left| 7 \right\rangle +\left| 55 \right\rangle  \right\}\notag\\
&{{H}_{2}}\left| 24 \right\rangle =\frac{-m}{2}\left\{ \left| 23 \right\rangle +\left| 22 \right\rangle +\left| 20 \right\rangle +\left| 32 \right\rangle +\left| 8 \right\rangle +\left| 56 \right\rangle  \right\}\notag\\
&{{H}_{2}}\left| 25 \right\rangle =\frac{-m}{2}\left\{ \left| 26 \right\rangle +\left| 27 \right\rangle +\left| 29 \right\rangle +\left| 17 \right\rangle +\left| 9 \right\rangle +\left| 57 \right\rangle  \right\}\notag\\
&{{H}_{2}}\left| 26 \right\rangle =\frac{-m}{2}\left\{ \left| 25 \right\rangle +\left| 28 \right\rangle +\left| 30 \right\rangle +\left| 18 \right\rangle +\left| 10 \right\rangle +\left| 58 \right\rangle  \right\}\notag\\
&{{H}_{2}}\left| 27 \right\rangle =\frac{-m}{2}\left\{ \left| 28 \right\rangle +\left| 25 \right\rangle +\left| 31 \right\rangle +\left| 19 \right\rangle +\left| 11 \right\rangle +\left| 59 \right\rangle  \right\}\notag\\
&{{H}_{2}}\left| 28 \right\rangle =\frac{-m}{2}\left\{ \left| 27 \right\rangle +\left| 26 \right\rangle +\left| 32 \right\rangle +\left| 20 \right\rangle +\left| 12 \right\rangle +\left| 60 \right\rangle  \right\}\notag\\
&{{H}_{2}}\left| 29 \right\rangle =\frac{-m}{2}\left\{ \left| 30 \right\rangle +\left| 31 \right\rangle +\left| 25 \right\rangle +\left| 21 \right\rangle +\left| 13 \right\rangle +\left| 61 \right\rangle  \right\}\notag\\
&{{H}_{2}}\left| 30 \right\rangle =\frac{-m}{2}\left\{ \left| 29 \right\rangle +\left| 32 \right\rangle +\left| 26 \right\rangle +\left| 22 \right\rangle +\left| 14 \right\rangle +\left| 62 \right\rangle  \right\}\notag
\end{align}
\begin{align}
&{{H}_{2}}\left| 31 \right\rangle =\frac{-m}{2}\left\{ \left| 32 \right\rangle +\left| 29 \right\rangle +\left| 27 \right\rangle +\left| 23 \right\rangle +\left| 15 \right\rangle +\left| 63 \right\rangle  \right\}\notag\\
&{{H}_{2}}\left| 32 \right\rangle =\frac{-m}{2}\left\{ \left| 31 \right\rangle +\left| 30 \right\rangle +\left| 28 \right\rangle +\left| 24 \right\rangle +\left| 16 \right\rangle +\left| 64 \right\rangle  \right\}\notag\\
&{{H}_{2}}\left| 33 \right\rangle =\frac{-m}{2}\left\{ \left| 34 \right\rangle +\left| 35 \right\rangle +\left| 37 \right\rangle +\left| 41 \right\rangle +\left| 49 \right\rangle +\left| 1 \right\rangle  \right\}\notag\\
&{{H}_{2}}\left| 34 \right\rangle =\frac{-m}{2}\left\{ \left| 33 \right\rangle +\left| 36 \right\rangle +\left| 38 \right\rangle +\left| 42 \right\rangle +\left| 50 \right\rangle +\left| 2 \right\rangle  \right\}\notag\\
&{{H}_{2}}\left| 35 \right\rangle =\frac{-m}{2}\left\{ \left| 36 \right\rangle +\left| 33 \right\rangle +\left| 39 \right\rangle +\left| 43 \right\rangle +\left| 51 \right\rangle +\left| 3 \right\rangle  \right\}\notag\\
&{{H}_{2}}\left| 36 \right\rangle =\frac{-m}{2}\left\{ \left| 35 \right\rangle +\left| 34 \right\rangle +\left| 40 \right\rangle +\left| 44 \right\rangle +\left| 52 \right\rangle +\left| 4 \right\rangle  \right\}\notag\\
&{{H}_{2}}\left| 37 \right\rangle =\frac{-m}{2}\left\{ \left| 38 \right\rangle +\left| 39 \right\rangle +\left| 33 \right\rangle +\left| 45 \right\rangle +\left| 53 \right\rangle +\left| 5 \right\rangle  \right\}\notag\\
&{{H}_{2}}\left| 38 \right\rangle =\frac{-m}{2}\left\{ \left| 37 \right\rangle +\left| 40 \right\rangle +\left| 34 \right\rangle +\left| 46 \right\rangle +\left| 54 \right\rangle +\left| 6 \right\rangle  \right\}\notag\\
&{{H}_{2}}\left| 39 \right\rangle =\frac{-m}{2}\left\{ \left| 40 \right\rangle +\left| 37 \right\rangle +\left| 35 \right\rangle +\left| 47 \right\rangle +\left| 55 \right\rangle +\left| 7 \right\rangle  \right\}\notag\\
&{{H}_{2}}\left| 40 \right\rangle =\frac{-m}{2}\left\{ \left| 39 \right\rangle +\left| 38 \right\rangle +\left| 36 \right\rangle +\left| 48 \right\rangle +\left| 56 \right\rangle +\left| 8 \right\rangle  \right\}\notag
\end{align}
\begin{align}
&{{H}_{2}}\left| 41 \right\rangle =\frac{-m}{2}\left\{ \left| 42 \right\rangle +\left| 43 \right\rangle +\left| 45 \right\rangle +\left| 49 \right\rangle +\left| 57 \right\rangle +\left| 9 \right\rangle  \right\}\notag\\
&{{H}_{2}}\left| 42 \right\rangle =\frac{-m}{2}\left\{ \left| 41 \right\rangle +\left| 44 \right\rangle +\left| 46 \right\rangle +\left| 34 \right\rangle +\left| 58 \right\rangle +\left| 10 \right\rangle  \right\}\notag\\
&{{H}_{2}}\left| 43 \right\rangle =\frac{-m}{2}\left\{ \left| 44 \right\rangle +\left| 41 \right\rangle +\left| 47 \right\rangle +\left| 35 \right\rangle +\left| 59 \right\rangle +\left| 11 \right\rangle  \right\}\notag\\
&{{H}_{2}}\left| 44 \right\rangle =\frac{-m}{2}\left\{ \left| 43 \right\rangle +\left| 42 \right\rangle +\left| 48 \right\rangle +\left| 36 \right\rangle +\left| 60 \right\rangle +\left| 12 \right\rangle  \right\}\notag\\
&{{H}_{2}}\left| 45\right\rangle =\frac{-m}{2}\left\{ \left| 46 \right\rangle +\left| 47 \right\rangle +\left| 41 \right\rangle +\left| 37 \right\rangle +\left| 61 \right\rangle +\left| 13 \right\rangle  \right\}\notag\\
&{{H}_{2}}\left| 46 \right\rangle =\frac{-m}{2}\left\{ \left| 45 \right\rangle +\left| 48 \right\rangle +\left| 42 \right\rangle +\left| 38 \right\rangle +\left| 62 \right\rangle +\left| 14 \right\rangle  \right\}\notag\\
&{{H}_{2}}\left| 47 \right\rangle =\frac{-m}{2}\left\{ \left| 48 \right\rangle +\left| 45 \right\rangle +\left| 43 \right\rangle +\left| 39 \right\rangle +\left| 63 \right\rangle +\left| 15 \right\rangle  \right\}\notag\\
&{{H}_{2}}\left| 48 \right\rangle =\frac{-m}{2}\left\{ \left| 47 \right\rangle +\left| 46 \right\rangle +\left| 44 \right\rangle +\left| 40 \right\rangle +\left| 64 \right\rangle +\left| 16 \right\rangle  \right\}\notag\\
&{{H}_{2}}\left| 49 \right\rangle =\frac{-m}{2}\left\{ \left| 50 \right\rangle +\left| 51 \right\rangle +\left| 53 \right\rangle +\left| 57 \right\rangle +\left| 33 \right\rangle +\left| 17 \right\rangle  \right\}\notag\\
&{{H}_{2}}\left| 50 \right\rangle =\frac{-m}{2}\left\{ \left| 49 \right\rangle +\left| 52 \right\rangle +\left| 54 \right\rangle +\left| 58 \right\rangle +\left| 34 \right\rangle +\left| 18 \right\rangle  \right\}\notag
\end{align}
\begin{align}
&{{H}_{2}}\left| 51 \right\rangle =\frac{-m}{2}\left\{ \left| 52 \right\rangle +\left| 49 \right\rangle +\left| 55 \right\rangle +\left| 59 \right\rangle +\left| 35 \right\rangle +\left| 19 \right\rangle  \right\}\notag\\
&{{H}_{2}}\left| 52 \right\rangle =\frac{-m}{2}\left\{ \left| 51 \right\rangle +\left| 50 \right\rangle +\left| 56 \right\rangle +\left| 60 \right\rangle +\left| 36 \right\rangle +\left| 20 \right\rangle  \right\}\notag\\
&{{H}_{2}}\left| 53 \right\rangle =\frac{-m}{2}\left\{ \left| 54 \right\rangle +\left| 55 \right\rangle +\left| 49 \right\rangle +\left| 61 \right\rangle +\left| 37 \right\rangle +\left| 21 \right\rangle  \right\}\notag\\
&{{H}_{2}}\left| 54 \right\rangle =\frac{-m}{2}\left\{ \left| 53 \right\rangle +\left| 56 \right\rangle +\left| 50 \right\rangle +\left| 62 \right\rangle +\left| 38 \right\rangle +\left| 22 \right\rangle  \right\}\notag\\
&{{H}_{2}}\left| 55 \right\rangle =\frac{-m}{2}\left\{ \left| 56 \right\rangle +\left| 53 \right\rangle +\left| 51 \right\rangle +\left| 63 \right\rangle +\left| 39 \right\rangle +\left| 23 \right\rangle  \right\}\notag\\
&{{H}_{2}}\left| 56 \right\rangle =\frac{-m}{2}\left\{ \left| 55 \right\rangle +\left| 54 \right\rangle +\left| 52 \right\rangle +\left| 64 \right\rangle +\left| 40 \right\rangle +\left| 24 \right\rangle  \right\}\notag\\
&{{H}_{2}}\left| 57 \right\rangle =\frac{-m}{2}\left\{ \left| 58 \right\rangle +\left| 59 \right\rangle +\left| 61 \right\rangle +\left| 49 \right\rangle +\left| 41 \right\rangle +\left| 25 \right\rangle  \right\}\notag\\
&{{H}_{2}}\left| 58 \right\rangle =\frac{-m}{2}\left\{ \left| 57 \right\rangle +\left| 60 \right\rangle +\left| 62 \right\rangle +\left| 50 \right\rangle +\left| 42 \right\rangle +\left| 26 \right\rangle  \right\}\notag\\
&{{H}_{2}}\left| 59 \right\rangle =\frac{-m}{2}\left\{ \left| 60 \right\rangle +\left| 57 \right\rangle +\left| 63 \right\rangle +\left| 51 \right\rangle +\left| 43 \right\rangle +\left| 27 \right\rangle  \right\}\notag\\
&{{H}_{2}}\left| 60 \right\rangle =\frac{-m}{2}\left\{ \left| 59 \right\rangle +\left| 58 \right\rangle +\left| 64 \right\rangle +\left| 52 \right\rangle +\left| 44 \right\rangle +\left| 28 \right\rangle  \right\}\notag\\
&{{H}_{2}}\left| 61 \right\rangle =\frac{-m}{2}\left\{ \left| 62 \right\rangle +\left| 63 \right\rangle +\left| 57 \right\rangle +\left| 53 \right\rangle +\left| 45 \right\rangle +\left| 29 \right\rangle  \right\}\notag\\
&{{H}_{2}}\left| 62 \right\rangle =\frac{-m}{2}\left\{ \left| 61 \right\rangle +\left| 64 \right\rangle +\left| 58 \right\rangle +\left| 54 \right\rangle +\left| 46 \right\rangle +\left| 30 \right\rangle  \right\}\notag\\
&{{H}_{2}}\left| 63 \right\rangle =\frac{-m}{2}\left\{ \left| 64 \right\rangle +\left| 61 \right\rangle +\left| 59 \right\rangle +\left| 55 \right\rangle +\left| 47 \right\rangle +\left| 31 \right\rangle  \right\}\notag\\
&{{H}_{2}}\left| 64 \right\rangle =\frac{-m}{2}\left\{ \left| 63 \right\rangle +\left| 62 \right\rangle +\left| 60 \right\rangle +\left| 56 \right\rangle +\left| 48 \right\rangle +\left| 32 \right\rangle  \right\}\notag\\
\end{align}
Finally the last term of the cluster Hamiltonian~\eqref{Hcluster.eqn} is the transverse field term,
\begin{align}
H_3=h\left\{\tau_0^z\tau_1^z+\tau_2^z+\tau_3^z+\tau_4^z+\tau_5^z\right\}
\end{align}
The effect of above term on the 64 bases is:
\begin{align}
&{{H}_{3}}\left| 1 \right\rangle =-6h\left| 1 \right\rangle,~~~~~~~~~~~~~~~~~~~~{{H}_{3}}\left| 2 \right\rangle =-4h\left| 2 \right\rangle,\notag\\
&{{H}_{3}}\left| 3 \right\rangle =-4h\left| 3 \right\rangle,~~~~~~~~~~~~~~~~~~~~ {{H}_{3}}\left| 4 \right\rangle =-2h\left| 4 \right\rangle,\notag\\
&{{H}_{3}}\left| 5 \right\rangle =-4h\left| 5 \right\rangle,~~~~~~~~~~~~~~~~~~~~{{H}_{3}}\left| 6 \right\rangle =-2h\left| 6 \right\rangle\notag\\
&{{H}_{3}}\left| 7 \right\rangle =-4h\left| 7 \right\rangle,~~~~~~~~~~~~~~~~~~~~{{H}_{3}}\left| 8 \right\rangle =0,\notag\\
&{{H}_{3}}\left| 9 \right\rangle =-4h\left| 9 \right\rangle,~~~~~~~~~~~~~~~~~~~~{{H}_{3}}\left| 10 \right\rangle =-2h\left| 10 \right\rangle,\notag\\
&{{H}_{3}}\left| 11 \right\rangle =-2h\left| 11 \right\rangle,~~~~~~~~~~~~~~~~{{H}_{3}}\left| 12 \right\rangle =0\notag\\
&{{H}_{3}}\left| 13 \right\rangle =-2h\left| 13 \right\rangle,~~~~~~~~~~~~~~~~{{H}_{3}}\left| 14 \right\rangle =0,\notag\\
&{{H}_{3}}\left| 15 \right\rangle =0,~~~~~~~~~~~~~~~~~~~~~~~~~~~~~{{H}_{3}}\left| 16 \right\rangle =2h\left| 16 \right\rangle,\notag\\
&{{H}_{3}}\left| 17 \right\rangle =-4h\left| 17 \right\rangle,~~~~~~~~~~~~~~~~{{H}_{3}}\left| 18 \right\rangle =-2h\left| 18 \right\rangle,\notag\\
&{{H}_{3}}\left| 19 \right\rangle =-2h\left| 19 \right\rangle,~~~~~~~~~~~~~~~~{{H}_{3}}\left| 20 \right\rangle =0,\notag\\
&{{H}_{3}}\left| 21 \right\rangle =-2h\left| 21 \right\rangle,~~~~~~~~~~~~~~~~{{H}_{3}}\left| 22 \right\rangle =0,\notag\\
&{{H}_{3}}\left| 23 \right\rangle =0,~~~~~~~~~~~~~~~~~~~~~~~~~~~~~{{H}_{3}}\left| 24 \right\rangle =2h\left| 24 \right\rangle\notag\\
&{{H}_{3}}\left| 25 \right\rangle =-2h\left| 25 \right\rangle,~~~~~~~~~~~~~~~~{{H}_{3}}\left| 26 \right\rangle =0,\notag\\
&{{H}_{3}}\left| 27 \right\rangle =0,~~~~~~~~~~~~~~~~~~~~~~~~~~~~~{{H}_{3}}\left| 28 \right\rangle =2h\left| 28 \right\rangle,\notag\\
&{{H}_{3}}\left| 29 \right\rangle =0,~~~~~~~~~~~~~~~~~~~~~~~~~~~~~{{H}_{3}}\left| 30 \right\rangle =2h\left| 30 \right\rangle\notag\\
&{{H}_{3}}\left| 31 \right\rangle =2h\left| 31 \right\rangle,~~~~~~~~~~~~~~~~~~~{{H}_{3}}\left| 32 \right\rangle =4h\left| 32\right\rangle,\notag\\
&{{H}_{3}}\left| 33 \right\rangle =-4h\left| 33 \right\rangle,~~~~~~~~~~~~~~~~{{H}_{3}}\left| 34 \right\rangle =-2h\left| 34 \right\rangle\notag\\
&{{H}_{3}}\left| 35 \right\rangle =-2h\left| 35 \right\rangle,~~~~~~~~~~~~~~~~{{H}_{3}}\left| 36 \right\rangle =0\notag\\
&{{H}_{3}}\left| 37 \right\rangle =-2h\left| 37 \right\rangle,~~~~~~~~~~~~~~~~{{H}_{3}}\left| 38 \right\rangle =0\notag\\
&{{H}_{3}}\left| 39 \right\rangle =0,~~~~~~~~~~~~~~~~~~~~~~~~~~~~~~{{H}_{3}}\left| 40 \right\rangle =2h\left| 40 \right\rangle,\notag\\
&{{H}_{3}}\left| 41 \right\rangle =-2h\left| 41 \right\rangle,~~~~~~~~~~~~~~~~{{H}_{3}}\left| 42 \right\rangle =0\notag\\
&{{H}_{3}}\left| 43 \right\rangle =0,~~~~~~~~~~~~~~~~~~~~~~~~~~~~~~{{H}_{3}}\left| 44 \right\rangle =2h\left| 44 \right\rangle,\notag\\
&{{H}_{3}}\left| 45 \right\rangle =0,~~~~~~~~~~~~~~~~~~~~~~~~~~~~~~{{H}_{3}}\left| 46 \right\rangle =2h\left| 46 \right\rangle,\notag\\
&{{H}_{3}}\left| 47 \right\rangle =2h\left| 47 \right\rangle,~~~~~~~~~~~~~~~~~~~{{H}_{3}}\left| 48 \right\rangle =4h\left| 48 \right\rangle\notag\\
&{{H}_{3}}\left| 49 \right\rangle =-2h\left| 49 \right\rangle,~~~~~~~~~~~~~~~~{{H}_{3}}\left| 50 \right\rangle =0,\notag\\
&{{H}_{3}}\left| 51 \right\rangle =0,~~~~~~~~~~~~~~~~~~~~~~~~~~~~~{{H}_{3}}\left| 52 \right\rangle =2h\left| 52 \right\rangle,\notag\\
&{{H}_{3}}\left| 53 \right\rangle =0,~~~~~~~~~~~~~~~~~~~~~~~~~~~~~{{H}_{3}}\left| 54 \right\rangle =2h\left| 54 \right\rangle,\notag\\
&{{H}_{3}}\left| 55 \right\rangle =2h\left| 55 \right\rangle,~~~~~~~~~~~~~~~~~~{{H}_{3}}\left| 56 \right\rangle =4h\left| 56 \right\rangle,\notag\\
&{{H}_{3}}\left| 57 \right\rangle =0,~~~~~~~~~~~~~~~~~~~~~~~~~~~~~{{H}_{3}}\left| 58 \right\rangle =2h\left| 58 \right\rangle,\notag\\
&{{H}_{3}}\left| 59 \right\rangle =2h\left| 59 \right\rangle,~~~~~~~~~~~~~~~~~~~{{H}_{3}}\left| 60 \right\rangle =4h\left| 60 \right\rangle,\notag\\
&{{H}_{3}}\left| 61 \right\rangle =2h\left| 61 \right\rangle,~~~~~~~~~~~~~~~~~~~{{H}_{3}}\left| 62 \right\rangle =4h\left| 62 \right\rangle,\notag\\
&{{H}_{3}}\left| 63 \right\rangle =4h\left| 63 \right\rangle,~~~~~~~~~~~~~~~~~~~{{H}_{3}}\left| 64 \right\rangle =6h\left| 64 \right\rangle.
\end{align}
The above $64\times 64$ Hamiltonian can be diagonalized on computer even without resort to
group theory methods.

\end{document}